\documentclass[lettersize,journal]{IEEEtran}
\usepackage{amsmath,amsfonts}
\usepackage{algorithmic}
\usepackage{algorithm}
\usepackage{array}
\usepackage[caption=false,font=normalsize,labelfont=sf,textfont=sf]{subfig}
\usepackage{textcomp}
\usepackage{stfloats}
\usepackage{url}
\usepackage{verbatim}
\usepackage{cite}
\usepackage{amsmath}
\usepackage{bbm}
\usepackage{float}
\usepackage{graphicx}
\usepackage{subcaption}
\usepackage{caption}
\usepackage{url}
\usepackage{booktabs}
\usepackage{pdflscape}
\usepackage{adjustbox}
\usepackage{lscape}
\usepackage{longtable}
\usepackage{tabularx}
\usepackage{comment}
\usepackage{bm}
\usepackage{xspace}
\usepackage{colortbl}
\usepackage{hyperref}
\usepackage{xcolor}
\usepackage{multicol, multirow}

\newcommand{\mypar}[1]{\noindent\textbf{#1}}
\newcommand{\mysubpar}[1]{\noindent\emph{\underline{#1}}}

\newcommand{\vct}[1]{\boldsymbol{#1}}

\newcommand{\exebench}{EXE-Bench\xspace}

\begin{document}

\title{EXE-Bench: Ranking the Tradeoffs of AI-based Windows Malware Detectors for Real-World Usability}

%
%
%

\author{Andrea Ponte\textsuperscript{\textdagger}, Daniel Gibert\textsuperscript{\textdagger}, Matous Kozak, Dmitrijs Trizna, Maura Pintor~\IEEEmembership{Member,~IEEE}, Battista Biggio~\IEEEmembership{Fellow,~IEEE}, Fabio Roli~\IEEEmembership{Fellow,~IEEE}, Luca Demetrio~\IEEEmembership{Member,~IEEE}

\thanks{Andrea Ponte, Luca Demetrio, Dmitrijs Trizna, and Fabio Roli are with the Department of
Informatics, Bioengineering, Robotics and Systems Engineering, University of
Genova, Italy (e-mail: andrea.ponte@edu.unige.it).}
\thanks{Daniel Gibert is with the Artificial Intelligence Research Institute (IIIA-CSIC), Spain (e-mail:daniel.gibert@iiia.csic.es).}
\thanks{Matous Kozak is with the Department of Information Security at Faculty of Information Technology, CTU in Prague, Czech Republic.}
\thanks{Dmitrijs Trizna is also with Department of Computer, Control and Management Engineering, Sapienza University, and AISLE.}
\thanks{Maura Pintor, Battista Biggio and also Fabio Roli are with the Department of Electrical and Electronic Engineering, University of Cagliari, Italy.}}

\maketitle

\begingroup
\renewcommand\thefootnote{\textdagger}
\footnotetext{These authors contributed equally to this work.}
\endgroup

\begin{abstract}
Due to the lack of systematic evaluations, we are not yet able to determine which AI-based Windows malware detector to deploy in production, since existing evaluations
(i) differ in terms of data used for both training and testing; 
(ii) do not consider temporal analysis to showcase whether models withstand the passage of time; 
(iii) avoid security evaluations with adversarial attacks that could highlight their brittleness against content-injection attacks; and (iv) neglect the computational requirements for deployment, risking slow inference on endpoints.
For these reasons, we develop EXE-Bench, a comprehensive benchmark of AI-based Windows malware detectors.
EXE-Bench assesses performance, temporal and adversarial robustness, and computational overhead, aggregating them into a single score for direct and fair model comparison.
Through EXE-Bench, we highlight how evaluations conducted only after deployment are suboptimal and unable to provide a complete picture of their performance.
In particular, through our analysis, we remark how much domain knowledge instilled through feature engineering is still extremely useful in this domain, resisting both time and adversarial attacks, in stark contrast with most of the deep networks that only excel right after deployment.
\end{abstract}

\begin{IEEEkeywords}
Malware Detection, Benchmark, Machine Learning, Deep Learning, Robustness, Drift.
\end{IEEEkeywords}

\section{Introduction}
Deciding which machine learning model to develop and deploy as a Windows malware detector is not an easy task, exacerbated by the fact that, during the last decade, we witnessed a surge in research proposals all claiming to achieve the best results.
In particular, in the context of static analysis, i.e., inferring maliciousness only by looking at the metadata and representation of programs without executing them, the predominant approaches can be divided into three distinct families: (i) feature-based~\cite{2018arXiv180404637A, 7413680, 10.1145/2857705.2857713}, that leverage domain-knowledge to extract an abstract, compact representation of programs used to train models; (ii) end-to-end~\cite{DBLP:conf/aaai/RaffBSBCN18,iclr_avastconv,8844623,GIBERT2021102159,10.1145/2016904.2016908,DBLP:journals/virology/GibertMPV19}, that train deep neural network directly on bytes, without pre-processing them; and (iii) certifiable~\cite{huang2023rsdel,gibert2023_randomizedsmoothing, 10.1145/3605764.3623914,saha2024drsm,10506708,gibert2024certifiedadversarialrobustnessmachine}, which leverage current findings in the image domain~\cite{DBLP:conf/icml/CohenRK19,NEURIPS2020_47ce0875} to produce models robust against minimal manipulations of the input, thus providing theoretical bounds on their security.

However, while all these models exhibit excellent performance in terms of accuracy and extremely-low false alarms, their evaluations cannot be directly compared due to completely different experimental settings that each of them applies.
In particular, all these models are trained on possibly different data sources, using different splits of the same datasets, or entirely different samples used for training and testing models.
Hence, while the results of those evaluations are consistent and valid within their scope, the comparison between techniques is not practical, leading to unfair settings.

Also, the main goal of these evaluations is solely confined to assessing accuracy on a single test set, neglecting that Windows malware detectors are deployed in an always-evolving environment subject to \emph{concept drift}~\cite{pendlebury2019tesseract}, i.e. the continuous change in the data distribution which violates the i.i.d. assumption considered when training models.

Measuring accuracy alone also hides fallacies of Windows malware detectors against \emph{adversarial EXEmples}~\cite{kolosnjaji2018adversarial, lucas2021malware, demetrio2021adversarial, demetrio2021functionality}, i.e. Windows malware minimally-perturbed to exploit the blind spots of machine learning techniques, thus evading detection.
Since most of the models were proposed before the advent of these attacks, they rightly did not consider these issues.
However, follow-up investigations incorporated adversarial EXEmples in their evaluations, but fell short in ways similar to those described earlier, even when proposing defenses~\cite{lucas2023adversarial, huang2023rsdel,gibert2023_randomizedsmoothing, 10.1145/3605764.3623914,saha2024drsm,10506708,gibert2024certifiedadversarialrobustnessmachine}.
Attacks are compared with different settings, thus providing partial results that cannot be directly compared with parallel studies.
Also, these evaluations neglect the fact that the security of machine learning models should be characterized against increasingly strong attacks, rather than testing only one configuration.
In particular, models that break with a few bytes alterations~\cite{demetrio2021functionality} highlight a stark sensitivity to input changes, making those techniques unreliable.

Lastly, all evaluations neglect the fact that these Windows malware detectors could be deployed in practice on endpoints, which are not likely to be equipped with the necessary hardware to support hundreds of predictions at once, or that they cannot redirect all their computational capabilities only to compute inference on a few downloaded programs.
Hence, while evaluations report the training time and resources needed to develop models, there is no clear way to compare those on the most frequent operation they are tasked with, which is prediction.
In practice, some models might be incredibly proficient in detecting malware, but computations take seconds or minutes, making them not usable in practice.

For these reasons, we overcome the assessment of accuracy alone, and we develop \exebench, a systematic benchmark to score models w.r.t. four main performance metrics (\autoref{sec:exebenchmark}), namely: (i) performance right after deployment, to assess their functionality in the present; (ii) performance over time, to assess whether their predictive capabilities remain stable even when facing the passage of time; (iii) performance against adversarial EXEmples, to assess whether they can resist attacks with increasing strength; and (iv) computational complexity, by assessing how fast they are when deployed.
All these metrics are combined into a single score providing a global ranking of each model, showcasing the ones achieving the best trade-off between these aspects for real-world usability.

All evaluations are conducted on the same identical settings (\autoref{sec:setup}), by training models on exactly the same data, the same training-test splits, with inference time computed on the same hardware.
To mimic the passage of time, we leverage a dataset collected 5 years later than the one used at training time, and models are evaluated on sequential 4-month windows, thus enabling a characterization of the degradation of performance in the presence of concept drift.
In this respect, while the other metrics are known from scattered previous work, the temporal metric we propose is novel, also pointing out possible fallacies of previous work~\cite{pendlebury2019tesseract} since they did not consider the number of samples contained in each quarter, wrongly influencing the temporal analysis.
Lastly, attacks are computed by selecting state-of-the-art techniques~\cite{demetrio2021adversarial, demetrio2021functionality} with increasingly stronger settings to characterize their robustness from negligible manipulations to heavier attacks injecting entire megabytes (MB) into malware programs, as done for other domains as well~\cite{biggio2018wild, cina2025attackbench}.

With \exebench we compare \textbf{30} different models, and we rank all of them according to our methodology (\autoref{sec:experimental_results}).
Our findings suggest that this domain still needs the use of domain-knowledge, instilled in the form of features extracted from Windows programs, as we highlight the superiority of those in almost all metrics.
While the rise of deep learning techniques rivals feature engineering on accuracy, most of them fall short in all the other metrics, being unable either to withstand the passage of time or remain robust against minimal perturbations.
Also, we are able to discourage certain technologies (like image classifiers retrofitted to perform malware detection) being completely suboptimal on most metrics, rising awareness of which are the reliable models to deploy.
As far as we know, this is the first fair and comprehensive benchmark in the domain of Windows malware detection with machine learning, and it can potentially stand out as a de-facto standard similar to previous approaches in the image domain~\cite{croce2021robustbench, cina2025attackbench}, paving the road towards safer, more reliable and more capable AI-based antivirus programs. We release all the code for computing the benchmark on GitHub,\footnote{\url{https://github.com/zangobot/exebenchmark}} along with an interactive dashboard hosting the collected results.\footnote{\url{https://exebench.github.io/}}

\section{Background and Related Work}
\label{sec:background}

\mypar{Windows PE File Format.}
\begin{figure}[t]
    \centering
    \includegraphics[width=0.85\linewidth]{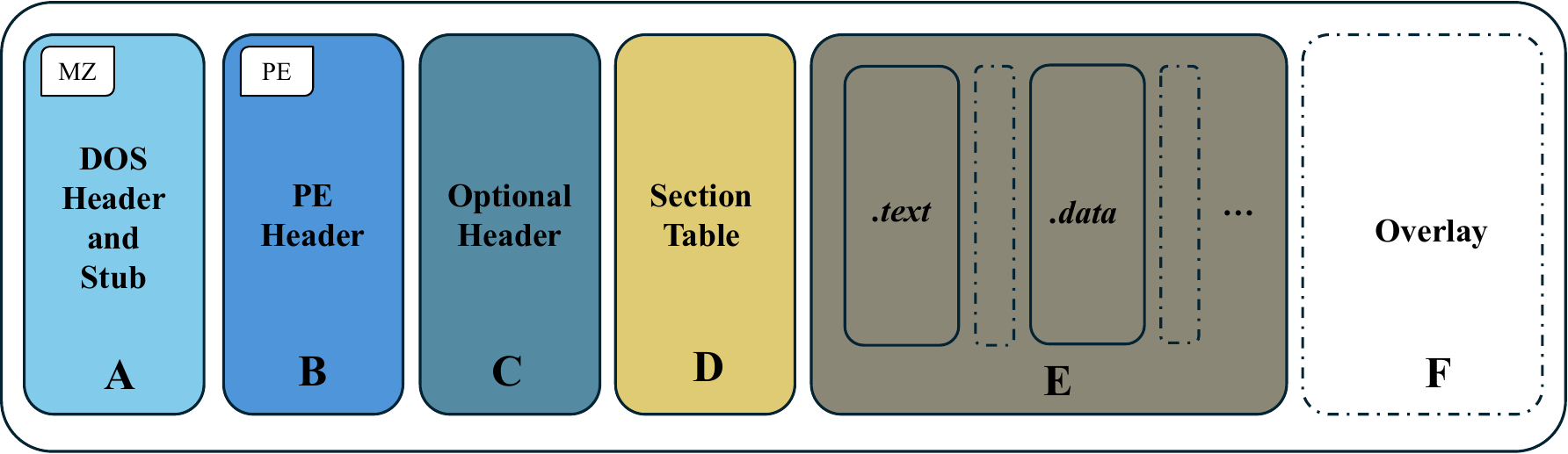}
    \caption{PE File Format.}
    \label{fig:pe}
\end{figure}
All AI-based antivirus programs need to digest Windows programs, stored on disk according to the Windows Portable Executable (PE) file format\footnote{\url{https://learn.microsoft.com/en-us/windows/win32/debug/pe-format}} as depicted in \autoref{fig:pe}.
Starting with the DOS Header and Stub (A) kept for retro-compatibility, each program holds its metadata inside the PE and Optional Headers (B,C), needed by the OS to correctly initialize the loading. 
These are followed by the Section Table (D), which instructs the OS where the main content of the program is physically located inside the file, i.e., the Sections (E).
Each section contains a different aspect of the program, like the compiled code, the resources, initialized values, etc.
Lastly, a program can have appended bytes at the end, i.e., the overlay (F), which is not considered by the loader.

\mypar{AI-based Windows Malware Detectors.}

\mysubpar{Feature-based models.}
This type of model relies on the manual extraction of hand-crafted features from PE files, used to train a machine learning classifier. Features can be extracted from the raw binary program~\cite{7413680,2018arXiv180404637A} or from its disassembled counterpart~\cite{10.1145/2857705.2857713}. 
However, disassembling a PE file is not always straightforward as it requires the use of tools such as Radare2,\footnote{\url{https://www.radare.org/n/}} IDA Pro,\footnote{\url{https://hex-rays.com/ida-pro}} or Ghidra,\footnote{\url{https://github.com/NationalSecurityAgency/ghidra}}, which add an additional pre-processing layer to the pipeline, increasing the processing time. 
Therefore, in this work we focus exclusively on the EMBER features~\cite{2018arXiv180404637A}, a well-known set of features extracted directly from the raw binary, including general file information, header metadata, section characteristics, imported and exported functions, and statistical features derived from bytes and strings.

\mysubpar{End-to-end models.} These architectures operate directly on raw byte sequences from executable files, allowing the models to learn hierarchical representations of low-level byte patterns without requiring feature engineering.
These are based on deep neural networks with different architectures, by either feeding bytes to an \emph{embedding layer}~\cite{DBLP:conf/aaai/RaffBSBCN18,iclr_avastconv,8844623,GIBERT2021102159}, i.e., a space learned at training time that imposes a distance metric over discrete values (bytes in this case), or processing each program as an image later fed to convolutions~\cite{10.1145/2016904.2016908,DBLP:journals/virology/GibertMPV19}.

\mysubpar{Certifiable models.}
This family of techniques~\cite{DBLP:conf/icml/CohenRK19,NEURIPS2020_47ce0875} has been proposed in the image domain to provide theoretical guarantees on robustness against adversarial attacks carried on with \emph{adversarial examples} --- minimally-perturbed test-time samples inducing classification errors.
Hence, these techniques have been also adapted to the Windows malware detection domain in the form of: (1) randomized smoothing~\cite{huang2023rsdel,gibert2023_randomizedsmoothing} and (2) de-randomized smoothing~\cite{10.1145/3605764.3623914,saha2024drsm,10506708,gibert2024certifiedadversarialrobustnessmachine}. 
Both are designed to provide theoretical guarantees on the robustness to small perturbations, differing in how the input is handled and how the certification is computed.
Randomized smoothing approaches transform a base non-robust classifier into a smoothed classifier that is probabilistically robust, by adding random noise to the input binary file multiple times and evaluating the classifier on each noisy instance. 
The final prediction is determined through majority voting, and it is possible to guarantee that small perturbations to the input will not alter the model's predictions. 
De-randomized smoothing~\cite{10.1145/3605764.3623914,saha2024drsm,10506708,gibert2024certifiedadversarialrobustnessmachine} approaches operate by partitioning an executable into fixed-sized chunks, which are then independently evaluated by the classifier.
Inference is then calculated through majority voting over the predictions of all chunks. 
Unlike randomized smoothing, these approaches can derive deterministic robustness certificates under different types of attacks if the number of predictions for the majority class exceeds that of the other class by a large enough margin~\cite{gibert2024certifiedadversarialrobustnessmachine}.

\mypar{Adversarial EXEmples.} 
Recent work highlighted the brittleness of AI-based Windows malware detectors when exposed to \emph{adversarial EXEmples}~\cite{kolosnjaji2018adversarial, demetrio2021adversarial, lucas2021malware}--- minimally-perturbed evasive malware samples.
Instead of being crafted by adding noise to each feature (i.e., computing attacks in the so-called \emph{feature space}), these attacks leverage functionality-preserving manipulations in the \emph{problem space}~\cite{pierazzi2020intriguing}, i.e., directly changing the programs without corrupting them.

\mypar{Related Work.}
Current evaluations of AI-based Windows malware detectors fall into two primary categories: (i)~\emph{algorithm performance benchmarks}, which compare various machine learning approaches including traditional methods (e.g., SVM, Random Forest) and modern neural networks (e.g., MalConv \cite{DBLP:conf/aaai/RaffBSBCN18}) across different public or private datasets~\cite{vinayakumar2019robust,fahim2025optimized}; and (ii)~\emph{adversarial robustness evaluations}, which assess model resilience against different evasion attacks~\cite{demetrio2021adversarial,imran2024evaluating,louthanova2024comparison,verwer2020robust}.
These studies evaluate both white-box (i.e., worst-case scenario, attackers own the models) and black-box (i.e., attackers can only access scores) attack scenarios, analyzing evasion rates, attack transferability across different target models, and perturbation characteristics. 
Research in this area ranges from small-scale controlled experiments to comprehensive multi-year evaluations, often incorporating structured comparison frameworks with specific attack constraints~\cite{demetrio2021adversarial,imran2024evaluating,louthanova2024comparison,verwer2020robust}.

\begin{figure*}[t]
\centering
\includegraphics[width=0.85\linewidth]{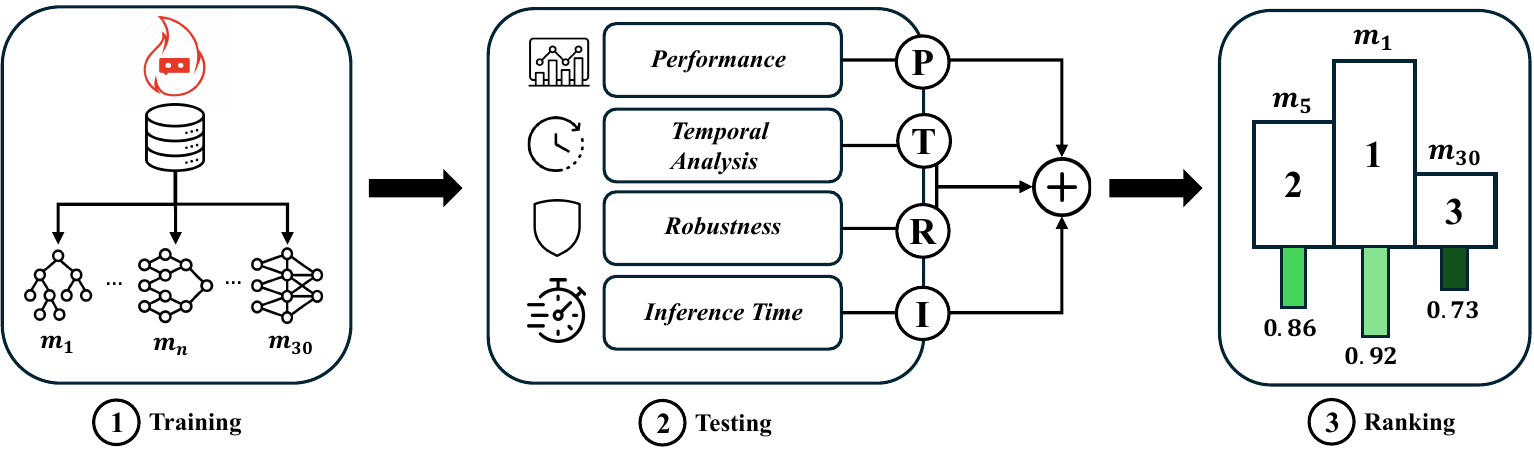}
\caption{Design of \exebench. Models are trained on the same data (1), evaluated through the lenses of test-time performance, stability over time, robustness against attacks, and inference time requirements (2). These results are aggregated into a single metric, used to rank models (3).}
\label{fig:benchmark-explanation}
\end{figure*}

\mysubpar{Limitations of Existing Evaluations.}
Current evaluations suffer from four critical issues that prevent reliable assessments and deployment decisions for AI-based Windows malware detectors.
First, \emph{dataset heterogeneity} creates incomparable results across studies. 
Previous evaluations use different data sources and scales, from small-scale assessments with 104 samples~\cite{demetrio2021adversarial} to larger datasets including thousands of samples~\cite{imran2024evaluating,fahim2025optimized}, thus making any comparison between findings impossible.
Thus, evaluations are conducted with different experimental setups leveraging different train-test splits, inconsistent choices of the detection thresholds, and mixed use of pretrained versus freshly trained models. 
Studies often evaluate different model architectures with hyperparameters picked with different policies, making performance comparisons meaningless and hindering reproducibility.
Second, \emph{temporal drift analysis} is largely absent from existing evaluations. 
Most works ignore the temporal nature of data rather than evaluating model degradation over time, despite malware and goodware evolution being a critical factor in production deployment. 
Studies using multi-year datasets~\cite{imran2024evaluating} typically aggregate results over periods without analyzing temporal stability patterns.
Third, \emph{robustness against attacks} is only partially evaluated, without constructing security evaluations~\cite{biggio2018wild} that assess the robustness against several adversarial attacks, from those that alter a few bytes to ones that inject portions of other programs into malicious samples.
As a result, models are only evaluated against specific strategies, rather than characterized as a whole.
Finally, current benchmarks focus on \emph{isolated metrics}, neglecting the fact that AI-based antivirus should achieve production-ready performance on many aspects such as computational requirements, deployment feasibility, and not only accuracy alone.

\section{\exebench: Systematic Evaluation of AI-based Windows Malware Detectors}
\label{sec:exebenchmark}
We now detail our methodology to rank machine learning-based Windows malware detectors, thus compiling the leaderboard of \exebench as shown in \autoref{fig:benchmark-explanation}.
All models are trained on the same data (step 1), and we proceed by computing four metrics that characterize the development and deployment of detectors (step 2): (i) the \emph{Performance Metric}, which quantifies the performance at deployment time; (ii) the \emph{Temporal Metric}, which quantifies the performance over time, after deployment; (iii) the \emph{Robustness Metric}, which quantifies the resistance against adversarial EXEmples; and (iv) the \emph{Inference Metric}, which quantifies the computational requirements when computing inference.
We then combine these into a single metric used to rank models (step 3), providing a general and model-agnostic leaderboard of all the considered models.

\mypar{Performance Metric}
We quantify the performance at deployment time through the application of the well-known F1 Score on the test set, depicted in this paper as $P(f) = F1(f)$.
In general, models are evaluated in terms of accuracy by fixing a detection threshold at 1\% False Positive Rate (FPR), thus being fair in terms of false alarms.
\textcolor{black}{Also, since models are trained and tested on the same data, this metric avoids biases caused by dataset heterogeneity (as described in \autoref{sec:background}}.
However, certifiable models cannot be tuned accordingly, since changing their output might interfere with their certification schema, and thus they are deployed to leverage majority voting only.
Since it is possible that also future extensions of the benchmark will pose a similar issue, we then rely on (i) calibrating models with a probability output at 1\%FPR, and (ii) keeping the certifiable models as is, to remain compliant with their development.
Hence, the F1 score is able to weight this slight advantage given to the certifiable models in our benchmark by also considering their non-fixed FPR.

\mypar{Temporal Metric.}
We propose a metric to quantify the performance that quantifies the performance in time, when the model is subject to concept drift~\cite{pendlebury2019tesseract} (i.e., the evolution of the distribution of programs induced by alterations of malware families, frameworks, class distributions on the collected data, and more).
To do so, we first need to define how the flow of time can be formalized in terms of future samples belonging to disjointed splits, i.e., sets containing samples attributed to a specific month of a specific year.
Let $\mathcal{D} = \{\mathcal{S}_1, \mathcal{S}_2, \ldots, \mathcal{S}_k\}$ be a dataset consisting of multiple splits, where each split $\mathcal{S}_i$ might contain a different number of samples.
Each split is defined as
$\mathcal{S}_i = \{(\vct x_{ij}, y_{ij})\}_{j=0}^{N_i}$, where $\vct x_{ij}$ is an input PE file, $y_{ij}$ its associated label (0 if benign, 1 if malicious), and $N_i$ the number of samples for split $i$. 
Thus, the temporal metric can be computed as:
\begin{equation}
    T(f) = \sum_i \frac{|S_i|}{\vert \mathcal{D} \vert} F1_{i}(f)
    \label{eq:temporal_metric}
\end{equation}
where $\mathcal{S}_i$ is a set of samples from a selected period of time (e.g., samples between January and March 2021), with cardinality $|S_i|$ used to weight the F1 score over the sum of all the samples in the dataset $|D|$.\textcolor{black}{We note here that the threshold used to compute the F1 score is the same that is used for the Performance metric.}
The metric has values between 0 and 1, with 1 meaning that the F1 score of the model was perfect in all temporal splits, while 0 means that the detector is failing on all splits.
We chose this metric instead of the AUC (Area Under the Curve), which was previously used for temporal evaluations~\cite{pendlebury2019tesseract} since the latter could provide incorrect results, as it does not take into account the number of samples used to compute each point of the curve (an intuitive example to support this choice is presented 
in \hyperref[appendix:auc]{Sect. A})

\mypar{Robustness Metric.} We devise a metric to express the robustness against adversarial EXEmples, inspired by previous work~\cite{cina2025attackbench}.
Similar to what is done for the image domain, we build a security evaluation curve, i.e., how the accuracy of the evaluated model changes when increasing the strength of the attacker, measured through a manipulation budget~\cite{biggio2018wild, pintor2021fast}.
In this domain, the budget of the attack can be expressed as the number of bytes that are either replaced or injected during the attack~\cite{kolosnjaji2018adversarial, demetrio2021adversarial, lucas2021malware}.
Hence, we define the \emph{Detection Rate at $\epsilon$} ($DR_\epsilon$) on a set of adversarial EXEmples $D_{EXE}$ as:
\begin{equation}
    DR_\epsilon(f) = \frac{1}{|D_{EXE}|}\sum_i \mathbbm{1}_{f(\vct x_i^\prime) = 1 \wedge d(\vct x_i, \vct x_i^\prime) \leq \epsilon}
\end{equation}
which counts how many adversarial EXEmples $\vct x^\prime \in D_{EXE}$ computed with a budget $\epsilon$ (w.r.t their original point $\vct x_i$) are still correctly classified as malicious.
The budget is bound to the Levenshtein distance (since we need to take into account both insertions and replacements) computed between $\vct x_i$ and $\vct x_i^\prime$.
This formulation does not depend on a specific manipulation or a specific optimization algorithm, thus it can be computed by leveraging different strategies all at once.
\textcolor{black}{Hence, this metric moves away from the robustness analysis performed in isolated settings, differently from previous work (as described in \autoref{sec:background}).}
Thus, we can quantify the degradation of robustness as the AUC, i.e., the integral of this curve:
\begin{equation}
    R(f) = \int DR_\epsilon(f) d\epsilon 
    \label{eq:robustness_metric}
\end{equation}
where, in theory, we sample all possible perturbation budgets.
Not feasible in practice, we approximate the integral by executing selected attacks, where $\epsilon$ depends on the used strategy (i.e., can be either fixed a priori, or estimate after the execution of the attack as later described in \autoref{sec:setup}).
Lastly, this metric can be normalized between 0 and 1 by dividing it by the area defined by the maximum $\epsilon$ estimated empirically (i.e., the maximum Levenshtein distance computed on all pairs of original and adversarial sample among all the selected attacks against all the models considered in the benchmark). 
Thus, when this metric has value 1, the evaluated model is perceived as robust, while 0 means the contrary.

\mypar{Inference Metric.}
The most frequent operation computed by a Windows malware detector is inference, hence it is necessary to rely not only on accurate but also fast models.
Also, it is very likely that these detectors could be deployed on general-purpose laptops with no GPU support. 
Hence, we track the average inference time $t_i(f)$ of the model $f$ on a set of training samples, and we quantify the computational requirements as:
\begin{equation}
    I(f) = e^{-t_i(f)}
    \label{eq:inference_metric}
\end{equation}
which penalizes models depending on how slow they are when computing predictions.
Thanks to the usage of the exponential with negative power, this metric has a theoretical maximum of 1 (with an impossible-to-achieve 0 seconds of inference time) and minimum of 0 (reached with incredibly slow models).\footnote{\textcolor{black}{We also tried a linear metric such as $ \frac{1}{1+t_i(f)}$, but it provided no meaningful changes in our results. We keep all the computations of this version of the metric on the public repository for transparency.}}

\mypar{\exebench leaderboards.}
Given these metrics, we can produce a summary score for each model, by combining the collected information in one single comprehensive metric:
\begin{equation}
    S(f) = \frac{P(f) + T(f) + R(f) + I(f)}{4}
    \label{eq:benchmark}
\end{equation}
This score is used to produce the leaderboard of \exebench, by ordering models w.r.t. this metric.

\section{Experimental Setup}
\label{sec:setup}
We now describe the datasets used during training and testing (\autoref{sec:data}), the considered machine learning models (\autoref{sec:models}), which adversarial attacks we use (\autoref{sec:advexe}), and how we quantify the computational requirements (\autoref{sec:computational}).
All the experiments have been conducted using the \texttt{maltorch} library,\footnote{\url{https://github.com/zangobot/maltorch}} which provided APIs for testing and computing adversarial attacks.

\subsection{Datasets and Temporal Analysis Setup}
\label{sec:data}
We leverage the EMBER 2017 dataset~\cite{2018arXiv180404637A} for training all models. 
Originally consisting of 400,000 benign, 400,000 malicious programs and 300,000 unlabeled programs, dated 2017 or earlier, we were able to retrieve 349,994/400,000 (87\%) benign and 399,992/400,000 (99\%) malicious programs using VirusTotal, by excluding the 300,000 unlabeled samples from our study.
The resulting dataset, consisting of 749,986 executable files, has been randomly divided into training (80\%), validation (10\%) and test (10\%) sets.

\begin{figure}[h]
    \centering
    \includegraphics[width=0.85\linewidth]{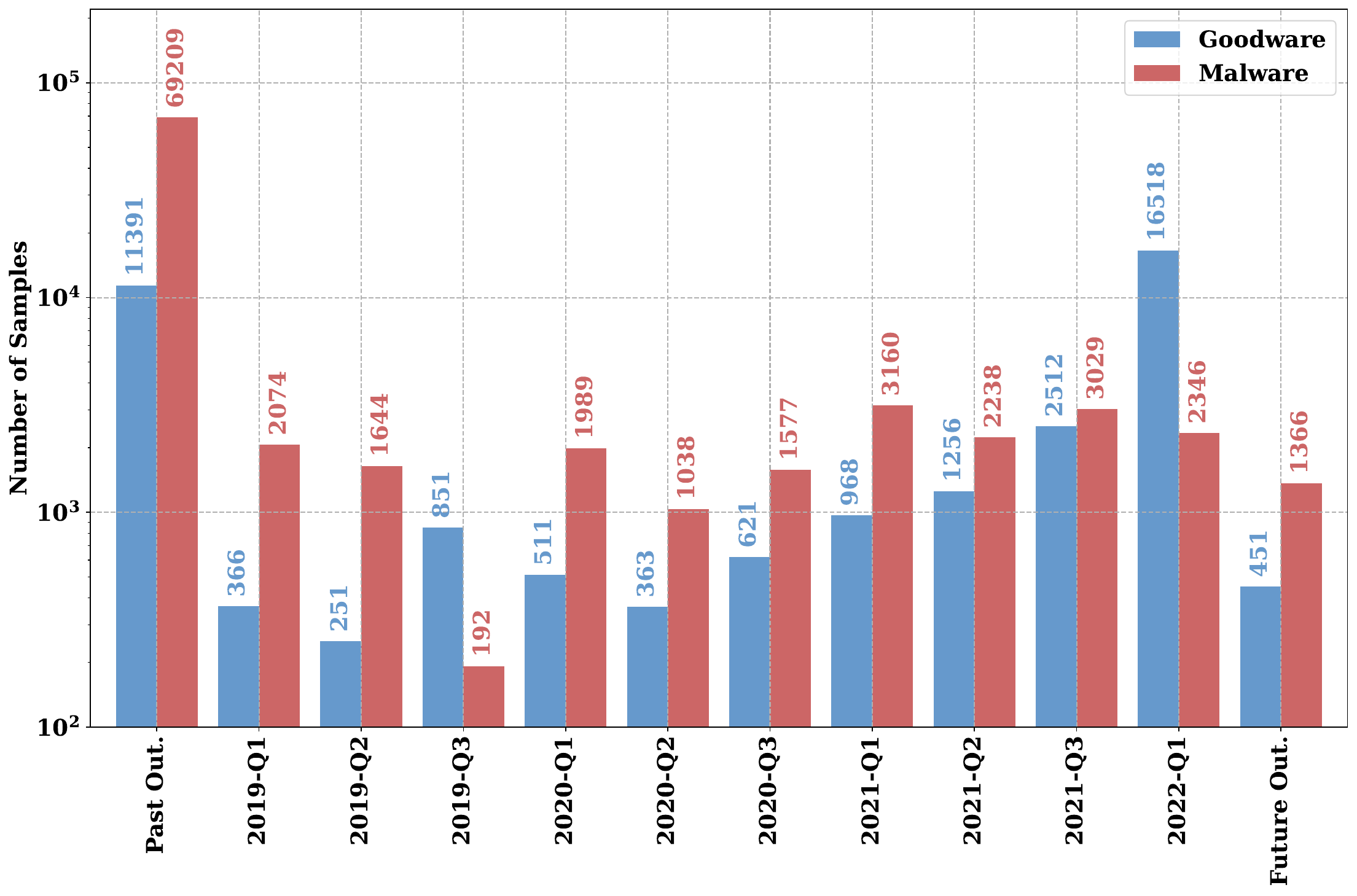}
    \caption{Temporal bins of Speakeasy Dataset in quarters, each with the quantities of malware and goodware for each.}
    \label{fig:timestamp_histogram}
\end{figure}

\mypar{Temporal Analysis.}
Since EMBER collects samples up to 2017, we use the Speakeasy dataset~\cite{trizna2022quo} to assess the capability of models over time.
The training set was collected in January 2022, while the test set was collected in April 2022. 
Since this dataset is only used to compute inference in this work, we merge the training and test sets, resulting in a total of 125,921 samples divided into 7 malware families.
To divide it into disjoint temporal bins, we use the timestamp contained in PE files, composing 10 bins from January 2019 (1 year after the most recent training data) to April 2022 (the latest period of gathering of Speakeasy), grouped in four-month periods.
Also, since timestamps could be easily manipulated, we discard from these datasets all values that are potentially invalid, namely \emph{past outliers} (containing distant past data and manipulated 0-epoch timestamps), and \emph{future outliers}, containing samples unrealistically dated in the future data. 
Similarly to the Android domain, where samples can be downloaded from AndroZoo or marketplaces reporting compilation timestamps~\cite{pendlebury2019tesseract}, in the Windows domain the ground truth on the first appearance can only be established through services like VirusTotal, since datasets like EMBER are only shared in this way.
\textcolor{black}{However, such information is not available for the Speakeasy dataset: at the time of writing, only 33\% of the programs used for the temporal analysis are also present on VirusTotal.
Moreover, only 7\% of those have a non-null ``first seen'' date (which should be the information to use as indicated by previous work~\cite{pendlebury2019tesseract}), forcing us to inspect the ``first submission date'' as an alternative (present in all reports).
However, 45\% of the considered samples have been submitted to VirusTotal after the collection date mentioned by the author of the dataset (January / April 2022 for training and test, respectively), thus making the acquired information unreliable (as shown in \autoref{tab:vt_timestamp_analysis}).}
Hence, our choice of relying on PE timestamps is sound and aligned with the proposed best practices~\cite{pendlebury2019tesseract}.
We report the quantities of malware and goodware for each bin in \autoref{fig:timestamp_histogram}.

\subsection{Models Selected for \exebench}
\label{sec:models}
Our benchmark comprises a total of \textbf{30} models, divided into \textit{feature-based}, \textit{end-to-end}, and \textit{certifiable} models.

\mypar{Feature-based Models.} We consider \emph{EMBER GBDT}, developed by Anderson et al.~\cite{2018arXiv180404637A}.
This model is implemented as a Gradient Boosting Decision Tree (GBDT)~\cite{friedman2001greedy} provided by the LightGBM library~\cite{ke2017lightgbm} and trained on the EMBER features.
\mypar{End-to-end Models.} We list and briefly describe here the end-to-end models we consider in our work.

\mysubpar{MalConv.} Developed by Raff et al.~\cite{DBLP:conf/aaai/RaffBSBCN18}, this model leverages a shallow convolutional neural network that takes as input sequences of 2,000,000  bytes ($\sim 2$ MB) and consists of an 8-dimensional embedding layer, a gated convolutional layer, followed by a global max-pooling and a feed-forward layer, for a total of 1,067,529 parameters.

\mysubpar{AvastConv.} Developed by Krčál et al.~\cite{iclr_avastconv}, this model takes as input sequences of 512,000 bytes ($\sim 500$ KB) and it is implemented through a convolutional neural network that consists of an 8-dimensional embedding layer, two convolutional layers, followed by a max-pooling and two convolutional layers, a global average pooling, and four feed-forward layers, with an overall parameter count of 904,697.

\mysubpar{BBDNN.} Developed by Coull et al.~\cite{8844623}, this model takes sequences of 102,400 bytes ($\sim 100$ KB) as input, and it is implemented as a convolutional neural network consisting of a 10-dimensional embedding layer, and five blocks of convolutional and a max-pooling layers.
Their output is passed to global max-pooling and global average-pooling layers concatenated along the feature dimension, culminating in a last feed-forward layer, for a total of 895,275 parameters.

\mysubpar{NGramConv.} Developed by Gibert et al.~\cite{GIBERT2021102159}, this model takes as input sequences of 512,000 ($\sim 500$ KB), approximating the extraction of n-grams through convolutions. 
This model leverages an embedding layer, a single convolutional layer with a small kernel size followed by a global max-pooling layer and a feed-forward layer, resulting in 17,957 parameters.

\mysubpar{ResNet18.} Originally proposed to deal with images~\cite{DBLP:conf/cvpr/HeZRS16},  we consider a ResNet18 (11,177,025 parameters) tailored to detect malware~\cite{10.1145/2016904.2016908,DBLP:journals/virology/GibertMPV19}, which takes in input programs as 256x256 grayscale images.
We consider this model as end-to-end since no domain knowledge is used to extract information.

\mypar{Certifiable Models.} We list here the certifiable models we consider in our work, divided into randomized and de-randomized smoothing as previously detailed in \autoref{sec:background}.

\mysubpar{Randomized Smoothing.} This family of models considers randomly-perturbed variations of input samples before computing predictions through majority voting.
Hence, they can be grouped into two categories based on the randomization strategy: (i) \emph{Byte Deletion}~\cite{huang2023rsdel}, which generates noisy instances by randomly deleting bytes; and (ii) \emph{Byte Ablation}~\cite{huang2023rsdel,gibert2023_randomizedsmoothing}, which generates noisy instances by randomly ablating bytes.
We thus produce one variant of the end-to-end models for each ablation technique, identified by the name of the model followed by RsDel for byte deletion or RS for byte ablation (e.g., MalConvRsDel and MalConvRS).

\mysubpar{De-Randomized Smoothing (DRS).} This family of models computes predictions by dividing samples into chunks, aggregating the decisions through majority voting.
They can be grouped into four categories: (i) \emph{Fixed-size DRS (FDRS)}~\cite{10.1145/3605764.3623914,gibert2024certifiedadversarialrobustnessmachine}, which splits the input into chunks of a predetermined size (e.g., 32768); (ii) \emph{K-Partitions DRS (KDRS)}~\cite{saha2024drsm}, which partitions the input into K partitions (e.g., K=12); (iii) \emph{Random DRS (RDRS)}~\cite{10506708}, which divides a file into chunks based on a configurable parameter (e.g., p=10\%) that determines the size of each chunk relative to the total file size, and it extracts multiple chunks from random locations; and (iv) \emph{Sequential DRS (SDRS)}~\cite{10506708}, determines the size of the chunks in the same way as RDRS but it extracts the chunks sequentially from the beginning of the file toward the end in a deterministic manner, without random sampling.
We thus produce one variant of the end-to-end models for each chunking technique, identified by the name of the model followed by the abbreviation of the respective chunking strategy (e.g., MalConvFDRS, MalConvKDRS, MalConvRDRS and MalConvSDRS).

\mypar{General Training Settings.}
The GBDT model has been implemented with LightGBM.
All neural networks (coded in PyTorch~\cite{paszke2019pytorch}) are trained with a batch size of 64, Adam optimizer with $10^{-3}$ as learning rate, the Binary Cross Entropy as loss, and an early stopping condition on the loss computed on a validation set, halting training whether such metric has not improved in the last 5 epochs (extended to 10 epochs for the certifiable models needed to take into account the effect of randomization and chunking, thus providing more stable results in terms of optimization).
We take into account the slight class imbalance by weighting them w.r.t. their proportion in our dataset.
All models have been trained and tested on a machine equipped with an Intel i9 14900KF (24 cores) CPU, 64 GB of RAM, and an Nvidia 4090 RTX GPU with 24GB of memory.

\subsection{Setup of Adversarial Attacks}
\label{sec:advexe}

\mypar{Data.}
To avoid possible corruptions due to obfuscated or packed samples, we selected 5000 unpacked programs from the EMBER test set for the robustness evaluation.
In this way, we are favoring the models in the benchmark to be tested against samples extracted from the time period in which they are more confident.
These samples have been selected by analyzing them with \emph{Detect it Easy},\footnote{\url{https://github.com/horsicq/Detect-It-Easy}} a well-known tool used for detecting packed samples, and keeping them whether the tool responded negatively.

\mypar{Adversarial Attacks.}
To provide insights on the robustness against adversarial attacks, we leverage three different techniques proposed in the literature~\cite{demetrio2021adversarial, demetrio2021functionality}.
To expand the applicability of our benchmark, we solely rely on black-box evaluations, i.e., strategies that only require the answers from the target to optimize attacks iteratively.
Thus, each of the attacks will leverage a genetic algorithm as an optimization algorithm, aligned with previous work~\cite{demetrio2021functionality}.
The latter works by generating a population of variants of the sample to optimize, and scores them according to the response of the target.
The process is repeated until the budget (i.e., the number of queries) is consumed.
All attacks hence use, at maximum, 500 queries, and produce 10 variants at each iteration of the attack.
Regarding the manipulation, these attacks leverage the following functionality-preserving manipulations that either re-write or inject new content into Windows programs.

\mysubpar{FullDOS.} 
This technique replaces the entire DOS header (A in \autoref{fig:pe}) with adversarial content~\cite{demetrio2021adversarial}, by keeping only the magic number \texttt{MZ} and the offset to the PE header (B in \autoref{fig:pe}).
Thus, the size of the manipulation is exactly the amount of bytes that are replaced by the attack.

\mysubpar{Content-shift.}
This technique injects content between the end of headers and the first section of the program (E in \autoref{fig:pe})~\cite{demetrio2021adversarial}.
To avoid corruptions, this content must preserve the \emph{file alignment}, i.e., a field in the Optional Header (C in \autoref{fig:pe}) specifying to the loader that the relevant content will be located at multiples of its value.
Hence, since different programs can have different alignment, we extract the maximum file alignment inside the dataset we sliced from our test samples, amounting to 4096 bytes.

\mysubpar{GAMMA.}
This technique harvests byte strings from benign programs and injects them into non-executable sections (D and E in \autoref{fig:pe}) of input samples~\cite{demetrio2021functionality}, thus relying on a pool of benign applications to use during its initialization step.
We setup two versions of this attack, by considering samples collected (i) from a fresh installation of Windows 11, to inject 5, 10, 20, 30, 50 sections of initialized data (\texttt{.rdata} section) from programs contained in \emph{sysWOW64}; and (ii) from the Speakeasy dataset, to inject 5, 10 sections of initialized data (\texttt{.rdata} section).
Since not all samples from Windows 11 contain the \texttt{.rdata} section, some attacks end up injecting empty sections with their section entry intact.
While this could be seen as a suboptimal setting, we include them into the analysis as they provide a valid perturbation that can be used to compute the metric (which, we remind, can contain all possible attacks regardless of their nature).
Lastly, GAMMA also uses a regularization parameter to provide a penalty term on the size of the perturbation during the optimization.
We set this parameter to 0 to inject as many bytes as possible.

\mypar{Transfer Evaluations.}
We also characterize robustness through \emph{transfer attacks}, i.e., adversarial EXEmples optimized against one model, and later tested on the real target~\cite{demontis2019adversarial}.
All the adversarial EXEmples that we compute, whether or not they evade the model used to optimize them, are saved and tested against all the other detectors.
This is especially useful since we were unable to compute attacks against certifiable models, due to their demanding computational requirements.
While this might favour certifiable models, we will later show that they do not occupy relevant places in the leaderboard.

\subsection{Computational Requirements}
\label{sec:computational}
Since inference time (pre-processing and forward operations) depends on the size of the input, we first sample the average ($\mu$) and standard deviation ($\sigma$) of the size of programs in the Speakeasy test set.
Then, we randomly sampled 500 malicious and 500 legitimate programs whose sizes fall within the interval $\mu \pm k\sigma$, fixing $k = 0.5$ to focus on inputs that are close to the center of their distribution, and we measured the average inference time on all models, including their pre-processing phase if any.
This measurement has been done on CPU, since models might be deployed on endpoints unlikely equipped with powerful GPUs to support frequent inference.

\begin{table}
\centering
\resizebox{\linewidth}{!}{
\begin{tabular}{lccccc|c}
\toprule
\textbf{Model} & \textbf{S} & \textbf{P} & \textbf{T} & \textbf{R} & \textbf{I} & \textbf{FR} \\
\midrule
\rowcolor{gray!10}
EmberGBDT & \textbf{0.86} & \textbf{0.99} & \textbf{0.78} & \textbf{0.80} & 0.87 & 4.25 \\
BBDnn & 0.85 & 0.97 & 0.67 & 0.79 & 0.98 & 4.75 \\
\rowcolor{gray!10}
BBDnnFDRS & 0.80 & 0.96 & 0.66 & 0.58 & \textbf{0.99} & 6.00 \\
BBDnnKDRS & 0.75 & 0.94 & 0.56 & 0.52 & \textbf{0.99} & 8.25 \\
\rowcolor{gray!10}
AvastStyleConvFDRS & 0.67 & 0.95 & 0.54 & 0.19 & \textbf{0.99} & 9.25 \\
AvastStyleConvKDRS & 0.67 & 0.93 & 0.57 & 0.19 & \textbf{0.99} & 10.00 \\
\rowcolor{gray!10}
NGramConv & 0.66 & 0.98 & 0.70 & 0.06 & 0.92 & 10.75 \\
BBDnnRsDel & 0.56 & 0.98 & 0.58 & 0.61 & 0.05 & 11.00 \\
\rowcolor{gray!10}
MalConv & 0.67 & \textbf{0.99} & 0.52 & 0.18 & 0.98 & 11.00 \\
NGramConvFDRS & 0.67 & 0.93 & 0.63 & 0.15 & 0.96 & 12.75 \\
\rowcolor{gray!10}
MalConvKDRS & 0.65 & 0.94 & 0.56 & 0.11 & \textbf{0.99} & 13.00 \\
BBDnnRDRS & 0.68 & 0.91 & 0.57 & 0.45 & 0.80 & 15.25 \\
\rowcolor{gray!10}
BBDnnSDRS & 0.69 & 0.91 & 0.57 & 0.50 & 0.79 & 15.25 \\
MalConvSDRS & 0.61 & 0.94 & 0.55 & 0.12 & 0.84 & 15.50 \\
\rowcolor{gray!10}
MalConvRsDel & 0.46 & 0.96 & 0.53 & 0.14 & 0.19 & 16.25 \\
MalConvRDRS & 0.61 & 0.94 & 0.56 & 0.11 & 0.84 & 16.25 \\
\rowcolor{gray!10}
AvastStyleConvRDRS & 0.61 & 0.94 & 0.50 & 0.16 & 0.82 & 16.50 \\
AvastStyleConvSDRS & 0.61 & 0.94 & 0.49 & 0.18 & 0.83 & 16.75 \\
\rowcolor{gray!10}
ResNet18 & 0.61 & 0.97 & 0.52 & 0.01 & 0.95 & 17.00 \\
NGramConvKDRS & 0.64 & 0.93 & 0.54 & 0.13 & 0.96 & 17.00 \\
\rowcolor{gray!10}
MalConvFDRS & 0.62 & 0.94 & 0.45 & 0.11 & \textbf{0.99} & 17.25 \\
NGramConvRsDel & 0.45 & 0.93 & 0.63 & 0.11 & 0.14 & 17.75 \\
\rowcolor{gray!10}
NGramConvSDRS & 0.56 & 0.93 & 0.54 & 0.11 & 0.66 & 19.00 \\
AvastStyleConvRsDel & 0.42 & 0.97 & 0.53 & 0.04 & 0.15 & 19.25 \\
\rowcolor{gray!10}
BBDnnRS & 0.47 & 0.67 & 0.52 & 0.62 & 0.09 & 20.25 \\
NGramConvRDRS & 0.56 & 0.93 & 0.54 & 0.11 & 0.65 & 21.25 \\
\rowcolor{gray!10}
AvastStyleConv & 0.45 & 0.78 & 0.03 & 0.01 & 0.98 & 23.50 \\
AvastStyleConvRS & 0.32 & 0.63 & 0.40 & 0.12 & 0.12 & 25.25 \\
\rowcolor{gray!10}
MalConvRS & 0.39 & 0.92 & 0.50 & 0.05 & 0.09 & 25.75 \\
NGramConvRS & 0.19 & 0.27 & 0.43 & 0.04 & 0.01 & 29.00 \\
\bottomrule
\end{tabular}}
\caption{\exebench leaderboard, reporting values of all metrics. We mark in bold the best values and ties.}
\label{tab:leaderboard}
\end{table}
\begin{figure*}
    \centering
    \includegraphics[width=0.97\linewidth]{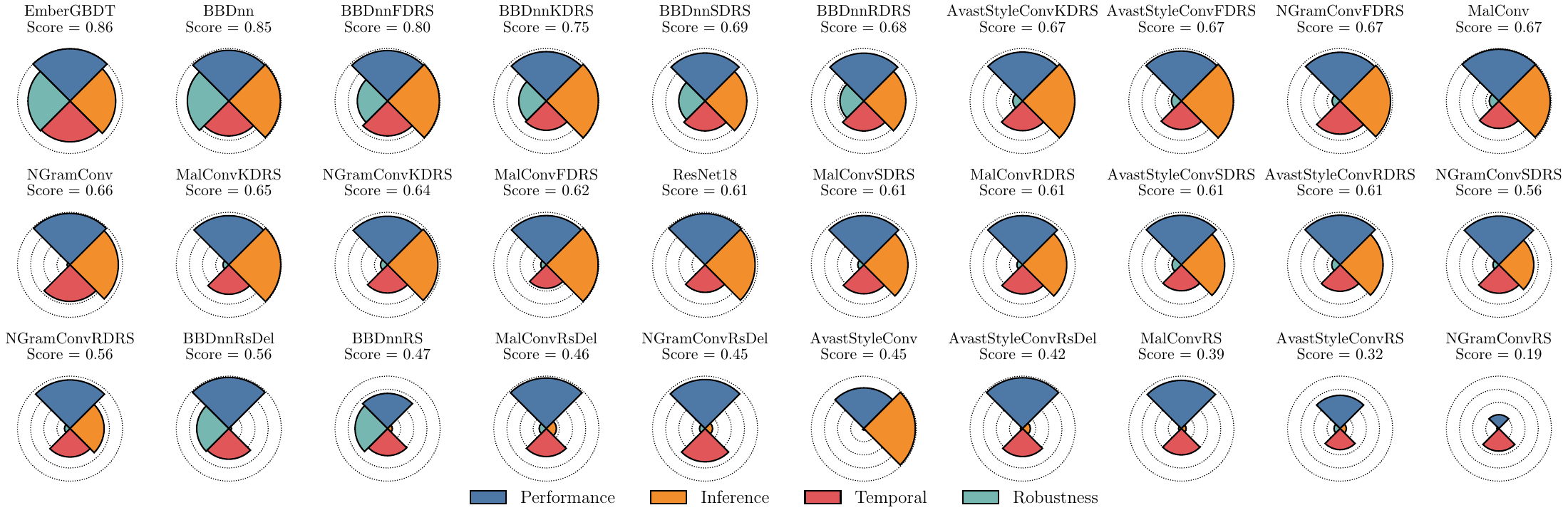}
    \caption{Radar plots summarizing the results of \exebench, ordering models by their score (S).}
    \label{fig:spie_benchmark}
\end{figure*}

\section{Experimental Results}
\label{sec:experimental_results}

\subsection{\exebench~Leaderboard}
\label{sec:exp_leaderboard}
We report in \autoref{tab:leaderboard} the results of our benchmark, displaying the values of our metrics for each model (also visually represented in \autoref{fig:spie_benchmark}).
\textcolor{black}{Through the lenses of the aggregated score (S), we order models from the top-performing to the least effective, allowing us to draw conclusions from the resulting leaderboard. 
This ranking is supported by a statistically significant level of agreement among the four evaluation metrics.
Specifically, Kendall's coefficient of concordance yields ($W=0.48, p=0.0021$), indicating moderate agreement between the metric-specific rankings. 
This result suggests that, while the four metrics capture complementary aspects of model behavior, they still produce sufficiently consistent orderings to justify the use of the aggregated score (S) as a meaningful summary of overall performance. 
For completeness, we also report the Friedman rank (FR) of each model, computed as the sum of its ranks across the individual metric leaderboards (P, T, R, and I), with lower values indicating better overall performance.
Thus, we can proclaim the GBDT model trained on EMBER features~\cite{2018arXiv180404637A} as the winner of our benchmark, followed by BBDnn~\cite{8844623}.}
This implies that the extraction of information through domain-knowledge has an impact on \emph{all} the relevant aspects of Windows malware detection, ranging from performance when deployed and in the future, to increased robustness to adversarial attacks.
This also confirms that, in this domain and contrary to image classification or detection and with this dataset size ($\sim 600k$ training samples), learning an abstract representation purely based on raw bytes cannot guarantee the preservation of all those aspects at once.

Interestingly, all certifiable models suffer from the aggregation of many metrics at once, showing the inevitable trade-off between accuracy on clean samples and robustness, exacerbated by the presence of drift in data distributions.
Among those, the de-randomized models that take into account chunk views on input programs (FDRS and KDRS) are the ones that balance the mentioned trade-off, while most of the other ones fail to reach the top of the leaderboard.
On the contrary, models trained with randomized smoothing techniques (RS and RsDel) are ranked the worst in our analysis.
While not peaking in accuracy-related metrics, these methods require each program to be sliced hundreds of times, expanding the computational cost to unbearable amounts, losing positions.

Both MalConv and BBDnn do not gain much from the introduction of certifiable techniques, thus performing better overall than their de-randomized and randomized smoothing versions.
On the other hand, both NGramConv and AvastStyleConv benefit from being trained with those techniques, as the certifiable variants appear higher in the leaderboard than the original version.
Thus, our benchmark also points out that applying certifiable approaches might not be a winning strategy by default, since there are many aspects that are impacted rather than the improvement of robustness alone.

\textcolor{black}{
\mypar{Sensitivity Analysis of Leaderboard.}
We now assess the stability of the leaderboard by performing a Leave-One-Out (LOO) sensitivity analysis, by recomputing the aggregated ranking after removing one metric at the time, measuring its correlation with the original ranking. 
We report in \autoref{tab:correlation} the Pearson's (r) and Spearman's ($\rho$) coefficients with their corresponding p-values.
High LOO correlations indicate that the ranking obtained without a given metric remains largely consistent with the original leaderboard, hence the removed dimension captures information already reflected by the remaining metrics. 
On the contrary, low correlations indicate that the omitted metric provides unique information to the benchmark, having thus a stronger influence on the final ranking.
Overall, the Inference metric yields the weakest LOO correlations ($\rho=0.2538$, $r=0.2623$), indicating that it has the largest impact on the leaderboard. 
In contrast, the Performance metric exhibits the highest agreement with the remaining metrics, suggesting a higher degree of redundancy with the other evaluation metrics.
This effect is also reflected in the magnitude of the ranking changes observed after removing each metric. 
The exclusion of the Inference metric produces the largest positional shifts across models: BBDnnRsDel moves from the 22nd to the 4th position, while removing the Temporal metric results in a maximum displacement of six positions. 
These findings suggest that the Inference metric captures aspects of model behavior that are not fully represented by the other evaluations, and therefore plays a key role in shaping the final leaderboard.
To facilitate further exploration of these effects, we developed and released an interactive dashboard\footnote{\url{https://exebench.github.io/}} that allows users to inspect the leaderboard and experiment with alternative metric weightings.} 
The latter can help practitioners in picking the best model that suit their need (e.g., limited hardware capacities, infrequent re-training, etc.) instead of using the average default weighting.

\subsection{Performance Metric Leaderboard}
\label{sec:exp_performance}
\begin{figure}
    \centering
    \includegraphics[width=\linewidth]{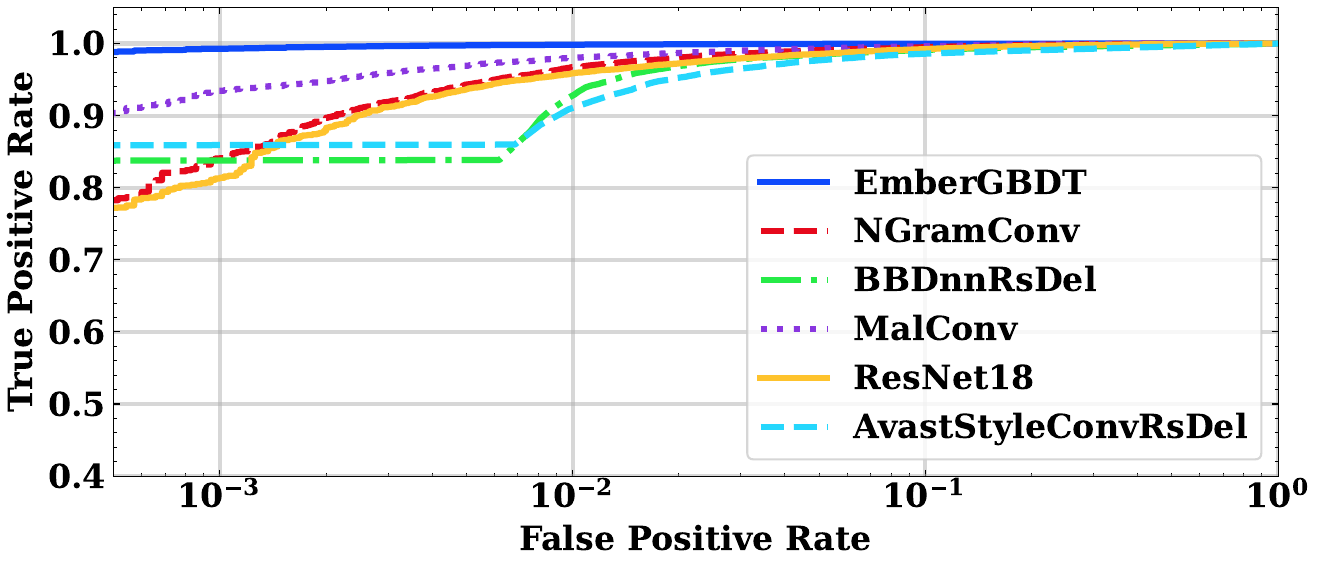}
    \caption{ROC curves of the top-six models according to the Performance metric.}
    \label{fig:roc_performance_metric}
\end{figure}

We report here the ranking based only on the Performance metric P, by listing the best six models in \autoref{tab:top-six} (second column), and also visualizing their ROC curve on the test set in \autoref{fig:roc_performance_metric}.
%
%
\begin{table*}[]
\centering
\begin{tabularx}{\textwidth}{@{}c|
  >{\raggedright\arraybackslash}Xc|
  >{\raggedright\arraybackslash}Xc|
  >{\raggedright\arraybackslash}Xc|
  >{\raggedright\arraybackslash}Xc@{}}
\toprule
\textbf{Rank} & \textbf{Model} & \textbf{P} 
             & \textbf{Model} & \textbf{T} 
             & \textbf{Model} & \textbf{R} 
             & \textbf{Model} & \textbf{I} \\ \midrule
\rowcolor{gray!10}\textbf{1} & EmberGBDT & 0.99 & EmberGBDT & 0.78 & EmberGBDT & 0.80 & AvastStyleConvKDRS & 0.99 \\
\textbf{2} & MalConv & 0.99 & NGramConv & 0.70 & BBDnn & 0.79 & BBDnnKDRS & 0.99 \\
\rowcolor{gray!10}\textbf{3} & NGramConv & 0.98 & BBDnn & 0.67 & BBDnnRS & 0.62 & AvastStyleConvFDRS & 0.99 \\
\textbf{4} & BBDnnRsDel & 0.98 & BBDnnFDRS & 0.66 & BBDnnRsDel & 0.61 & MalConvKDRS & 0.99 \\
\rowcolor{gray!10}\textbf{5} & ResNet18 & 0.97 & NGramConvFDRS & 0.63 & BBDnnFDRS & 0.58 & MalConvFDRS & 0.99 \\
\textbf{6} & AvastStyleConvRsDel & 0.97 & NGramConvRsDel & 0.63 & BBDnnKDRS & 0.52 & BBDnnFDRS & 0.99 \\
\bottomrule
\end{tabularx}
\caption{Top six models according to each metric of \exebench, displayed column-wise.}
\label{tab:top-six}
\end{table*}
When considering the performance on the test set alone, most of the regular versions of models are present among the top-performing ones, missing only AvastStyleConv (which achieved a suboptimal F1 score of 0.76 on the test set).
Interestingly, the RsDel certification schema (which generally decreases the ranking of models to which it is applied, as highlighted in \autoref{sec:exp_leaderboard}) slightly improves the F1 of BBDnn, being an incredible boost for AvastStyleConv (i.e., it improves F1 by $\sim0.2$).
By looking only at this leaderboard, the positions are different w.r.t. the complete one, as MalConv is ranked higher and also surpasses the second-best model BBDnn.
This is not surprising, since these models have been developed solely to excel in terms of accuracy, thus possibly neglecting the other metrics that we observe.
For instance, MalConv is not able to keep up with both the passage of time and robustness, falling down in the overall ranking as it is likely overfitting the distribution of the present data available during deployment.
On the contrary, as highlighted in \autoref{fig:roc_performance_metric}, no model is a match for the EmberGBDT, which achieves almost perfect results already at $10^{-3}$ FPR.
To summarize, this reduced leaderboard represents a breakdown of the regular comparisons between models, by just picking the ones with the best score in terms of detection rate and low false alarms.
However, this metric alone does not provide a full picture of the behavior of the analyzed models.
\begin{figure}
    \centering
    \includegraphics[width=\linewidth]{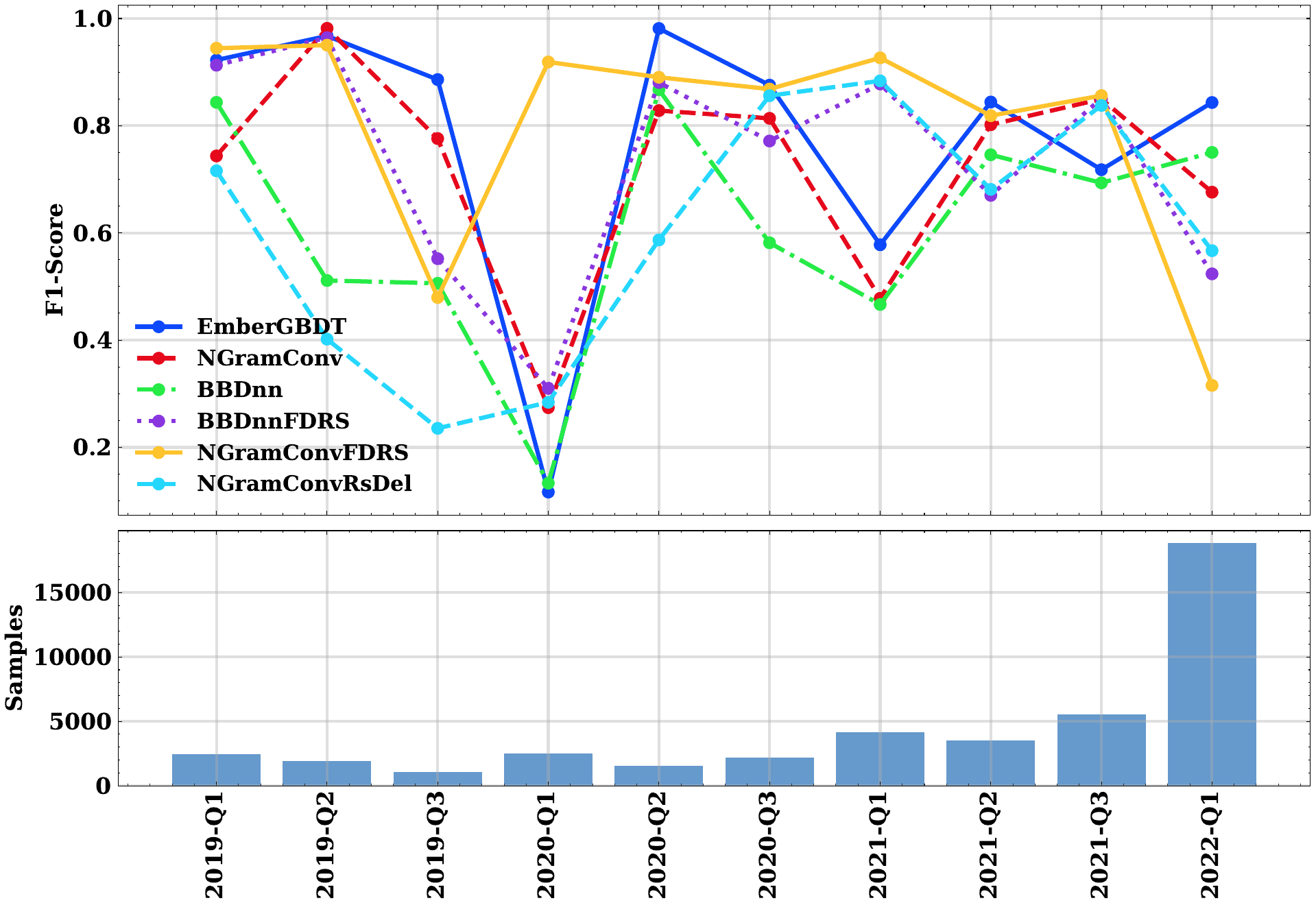}
    \caption{Fluctuations of the F1 score of the top six models, ranked by the temporal metric, and the number of samples contained in each temporal bin.}
    \label{fig:temporal_plot}
\end{figure}
\subsection{Temporal Metric Leaderboard}
\label{sec:exp_temporal}
We now report the ranking based on the Temporal metric T, by listing the best six models in \autoref{tab:top-six} (third column), while also depicting the fluctuations of the F1 score on the various bins in \autoref{fig:temporal_plot}.
We notice the presence of certifiable approaches among the best models, with de-randomized smoothing leading the way.
This could be explained by the fact that, while portions of code and data might change over time (i.e., being completely different in terms of bytes), the information contained inside headers (including the Import Address Table, that explains to the operating system which API to import) is likely to be less prone to such a change.
Hence, the FDRS methods, which segment each input program into chunks of fixed size, combined with the majority voting, enable these models to weigh more the decision on headers than the bytes contained in sections (which might be noisier from one bin to the other).
Interestingly (and differently from the analysis on the Android domain~\cite{pendlebury2019tesseract}), the top six models are not really strictly losing performance over time, fluctuating bin by bin.
These fluctuations can be caused by a mixture of factors, ranging from the inaccurate attribution due to timestamp manipulation, or by the scarcity of the number of samples contained in those bins.
However, in the context of a benchmark, the evaluation is fair for all models, since they are evaluated on the same data, by addressing the size of bins through the weighted average.
To summarize, this reduced leaderboard gives clear insights on which model can be more resistant to the passage of time, providing guidance on what to deploy in production.

\subsection{Robustness Metric Leaderboard}
\label{sec:exp_robustness}
\begin{figure}[t]
    \centering
    \includegraphics[width=0.9\linewidth]{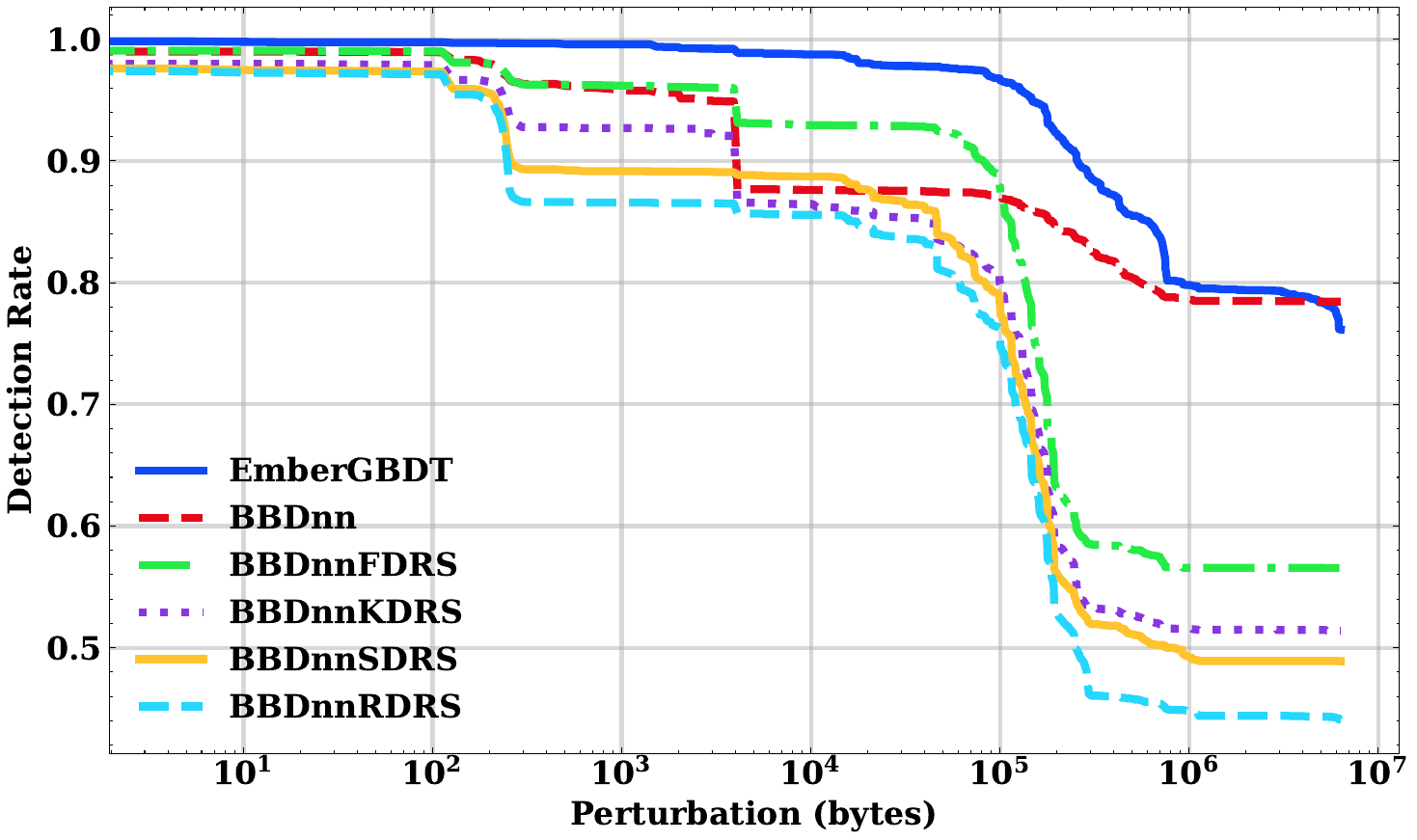}
    \caption{Security evaluations of the top-six models in terms of robustness.
    The x axis is the number of perturbed bytes, while the y axis is the Detection Rate (DR).}
    \label{fig:robustness_dr_eps}
\end{figure}
We now report the ranking based only on the Robustness metric R, by listing the best six models in \autoref{tab:top-six} (fourth column), while also depicting their security evaluation curves in \autoref{fig:robustness_dr_eps}.
GBDT and BBDnn obtain almost the same score, with GBDT being slightly better: as shown in \autoref{fig:robustness_dr_eps}, the neural network exhibits better robustness only after having injected a few MB into samples (far right of the figure).
However, except for BBDnn and its variants, all the other end-to-end models perform poorly.
In particular, upon inspection, we highlight that ResNet18, the image-based classification method, is broken against the weakest attacks of our benchmark, i.e., FullDOS and Content-shift.
This can be easily explained by the fact that folding programs into images (i.e., breaking byte strings into rows and columns) does not provide any spatial advantage, since different rows in the produced image are not necessarily semantically close.
Also, adversarial content is likely to form contiguous blocks that disrupt the convolution operations, easily misleading all the predictions.
Interestingly, the regular BBDnn model is more robust than its randomized and de-randomized smoothing versions with a striking gap in terms of the computed metric.
This could be caused by GAMMA injecting portions of legitimate programs into malware, used during the slicing provided by the randomization.
In particular, each of these methods computes inference only on a smaller view of the sample, which can now also contain bytes harvested from goodware programs.
Thus, many of the blocks are patterns of bytes considered legitimate, altering the decision of the majority voting schema.
Lastly, the robustness of BBDnn might also be an effect of its reduced input window (100KB), differently from all the other networks.
Hence, content-injection attacks like GAMMA might create plenty of sections whose content falls outside such input window, thus enlarging the perturbation is not bringing an advantage to the attacker.
To summarize, extracting features improves robustness against adversarial attacks, followed by neural networks that compute predictions by looking at a reduced portion of the entire input executable.
On the contrary, methods claimed as robust are not really effective against content-injection attacks, with randomization and majority voting system being their limit.

\subsection{Inference Metric Leaderboard}
\label{sec:exp_inference}
\begin{figure}[t]
    \centering
    \includegraphics[width=\linewidth]{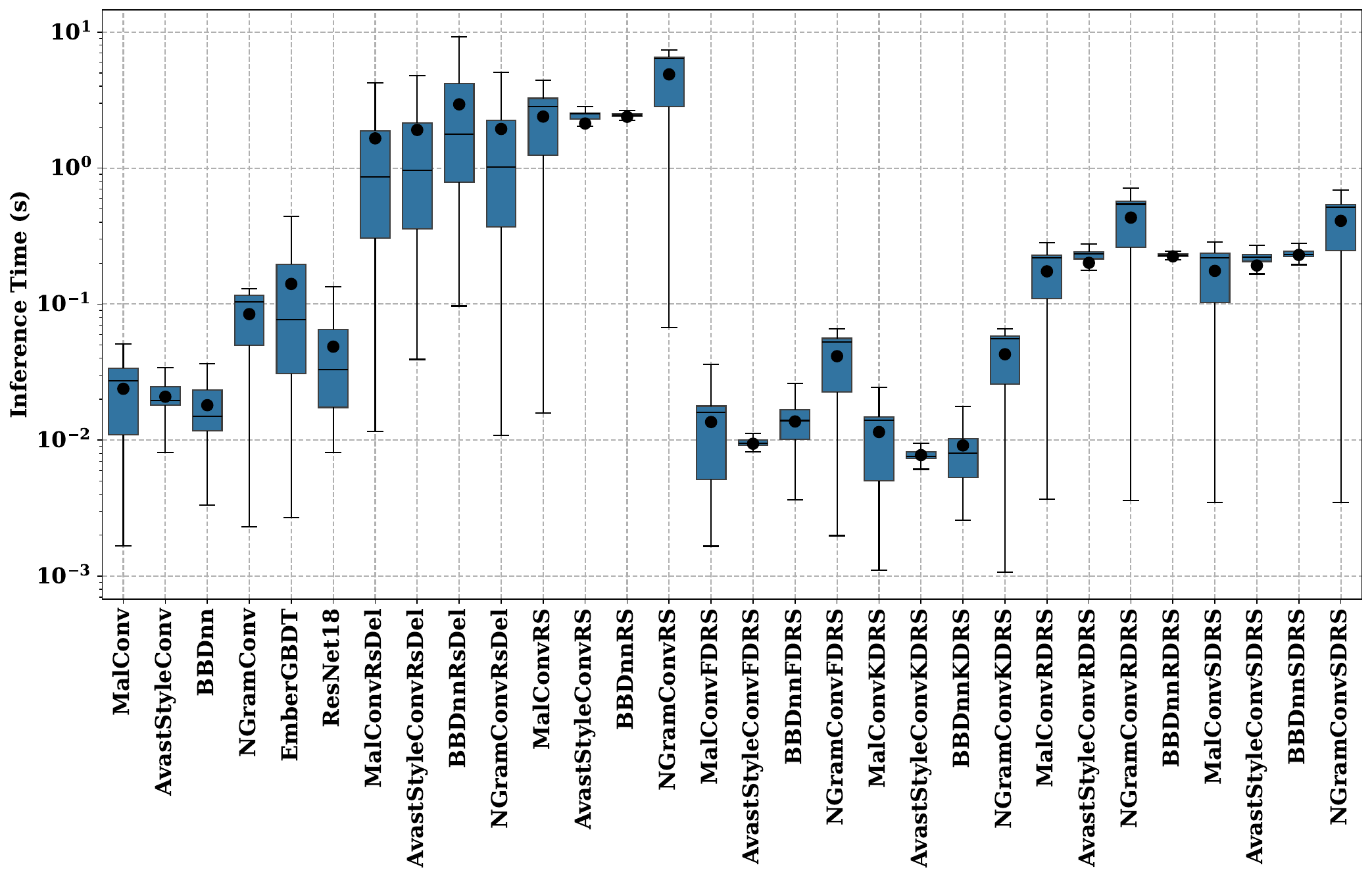}
    \caption{Boxplot with average (dot), median (line), and standard deviation (blue area) of inference time of all the models in the benchmark.}
    \label{fig:inference_times}
\end{figure}
We report the ranking based only on inference time by listing the six best models in the last column of \autoref{tab:top-six}, while also providing a comprehensive view on inference time expressed in seconds in \autoref{fig:inference_times}. Lastly, even if omitted from the benchmark, we report in \autoref{tab:training_time} the hours needed for training all models for completion.
Counter-intuitively, the fastest models are the de-randomized ones, surpassing all others in the reduced leaderboard.
Even if inference is computed multiple times due to the certification schema (e.g., 12 for KDRS), these models receive in input samples negligible in size, allowing very-fast predictions.
In particular, these are even faster than computing a prediction on the whole sample, as globally shown in \autoref{fig:inference_times}.
On the contrary, the slowest models are the ones using randomized smoothing, since inference is repeated 100 times on truncated samples, providing computational overhead required by the sampling.
To summarize, this reduced leaderboard provides a snapshot of which model can produce fast predictions.
While this metric alone cannot provide a solid ranking, it might be useful to prioritize speed w.r.t. other metrics, i.e., in contexts in which quick filters are needed before more rigorous analyses. 

\subsection{Concluding Summary}
\label{sec:exp_summary}

\mypar{Accuracy alone is not enough.} Our results show that being accurate on a test set is not the only aspect at deployment time. 
When complemented with efficiency, temporal and adversarial robustness, practitioners can collect a complete snapshot of the capabilities of models before deployment.

\mypar{Feature engineering makes the difference.} 
EMBER GBDT provides top rankings on all metrics except inference time, due to the inevitable feature extraction phase.
However, this model is still capable of computing inference in less than one second on a compact set of features, without relying on deploying a large convolutional neural network with less accurate results (but slightly faster).
Such a top-tier result underlines the idea that we still indeed need feature engineering in the domain of Windows malware detection, as we have not yet reached the stage at which deep neural networks surpass the application of domain knowledge.
Also, the application of domain knowledge also improves stability over time and robustness against attacks, making them a perfect fit for production environments.
These models can be rapidly trained (4 minutes) once all features have been processed, and updates can be shipped fast since all the already-collected samples are already processed, needing to rely on feature extraction of only the fresh ones (reasonably less than the pre-owned ones).

\mypar{Efficiency of De-randomized smoothing.} Even if this family of methods is not scoring a top rank in the leaderboard, it is interesting to notice that they are fast to train and deploy.
With more research, it is possible that they will conquer higher places in the ranking, while being more deployable-friendly.

\mypar{Unreliable image-based detection.} While classifying malware as an image might speed up the process, using images as a representation for Windows programs is neither robust to adversarial attacks nor robust to the passage of time, being among the worst w.r.t. these two aspects.
Hence, we deeply discourage the usage of these types of detectors in production.

\section{Limitations}
While our methodology is able to provide a clear winner by analyzing different axes, we acknowledge the presence of different limitations of \exebench, by also showing how these can be addressed and tackled with ease.

\mypar{Static analysis only.} All the models we consider only perform static analysis on samples, ignoring models trained on the output of dynamic analysis.
While the literature on the matter is vast~\cite{gibert2020rise}, these methods are characterized by huge computational and deployment costs due to the instantiation of emulators or sandboxes to safely detonate malware and track their behavior.
While our current analysis does not cover this aspect, \exebench is general enough to compute these metrics for any Windows detector.
This is achieved thanks to the generality of the metrics, since they all characterize a relevant aspect of models, regardless of their nature.

\mypar{EMBER dataset.} While the dataset we used is relevant, more recent data sources have been released, including SOREL~\cite{harang2020sorel}, and EMBER2024 version~\cite{joyce2025ember2024}.
While timely, their usage does not change the proposed methodology, and can be easily included in \exebench either as training data or future time splits without changing the overall methodology.

\mypar{Missing models.} While we evaluated state-of-the-art techniques, we have left behind some techniques leveraging other pre-processing or hardening approaches. 
We excluded transformer-based detectors due to the GPU memory required for training: prior work~\cite{kurlandski2026} reports comparable detection performance to our CNN-based models, but at substantially higher computational cost, with inference times 570.57\% higher than MalConv.
We also did not include methods based on the assembly representation of programs~\cite{10.1145/2857705.2857713, 10145856}, such as GNN-based approaches that operate on Control Flow Graph or Function Call Graph representations of programs, because many programs are packed or obfuscated.
In such cases, disassemblers may produce incomplete, misleading, or unreadable output, and the quality of their output highly depends on the source of these technologies.
Unfortunately, the best disassemblers are commercial products with expensive licenses, like IDA Pro,\footnote{\url{https://hex-rays.com/ida-pro}} which also introduce latency when computing inference times.
On the same note, we did not include adversarial training~\cite{lucas2023adversarial,10.1145/3658644.3690208, kozak2025updating} since it is extremely time-consuming and computationally expensive. 
For example, the fastest method reported in previous work~\cite{lucas2023adversarial,10.1145/3658644.3690208} required 5.7 days of training for just 3 epochs on fewer than 300k samples, while the slowest approach took months.
In contrast, our training dataset is twice their size, thus using this technique on more than 3 epochs would have required prohibitively long runtimes. 
Furthermore, the attacks used to compute adversarial EXEmples while training~\cite{lucas2023adversarial,10.1145/3658644.3690208} are either not publicly released or are missing components, making exact reproduction impractical.
Also, even if missing, the benchmark is still general enough to tackle the future inclusion of all AI-based Windows malware detectors without loss of generality.
The quantities we measure are model agnostic, and they solely depend on the performance and computational requirements of those models.

\mypar{Limitations of randomized-smoothing approaches.}
While randomized smoothing provides bounds on the accuracy when manipulating up to a specific number of bytes~\cite{huang2023rsdel} (i.e. the certified accuracy), we were unable to include such metric in our benchmark due to computational requirements.
For instance, RsDel methods would need $\sim4000$ randomizations of each sample to estimate the certified accuracy.
\textcolor{black}{
Moreover, during inference, the original implementations of RsDel and RS do not fit the VRAM of our GPUs, forcing us to cut samples according to the input window of each model.}
Nevertheless, these issues do not pose a problem to the benchmark itself, but rather highlight the need for more computational power to handle certifiable models.
In fact, without these simplifications, certifiable models would have sunk even lower in the benchmark due to their infeasible computational time.
The same can be said for their adversarial evaluation: directly attacking them would have required much more computational capacity.
Even so, the collected results highlight the necessity for more research in the context of certification.

\mypar{Biases of the temporal analysis.} \textcolor{black}{While we proved that timestamps are the only available information we can leverage, we acknowledge that the temporal evaluation might be biased by other factors, like the presence of timestomping techniques (i.e., misleading the temporal attribution by altering the original compilation timestamp\footnote{\url{https://attack.mitre.org/techniques/T1070/006/}}), by the specific era we have considered (2019-2022), or also by the vendor that originally acquired the dataset.
Nevertheless, we have followed the best practices streamlined by previous work~\cite{pendlebury2019tesseract}, and, whether it would be possible to retrieve the ground truth of the first seen dates, our leaderboard can be re-evaluated accordingly, as the way \exebench operates remains unchanged.}

\mypar{Missing attacks.} While we computed a massive amount of adversarial EXEmples, the literature contains other techniques which we do not have included into \exebench~\cite{lucas2021malware, song2022mab}.
In particular, some models might be weaker to some techniques and not others, or their input windows might be too small and thus cut all the injected adversarial content (e.g., BBDnn), thus providing an advantage in terms of robustness.
However, while these techniques require commercial tools like IDA Pro~\cite{lucas2021malware}, or are known to corrupt samples~\cite{song2022mab}, our methodology remains sound, and those attacks can be added at any time after having provided a correct implementation of those techniques.

\mypar{Absence of metrics for Malware Classification.} 
We did not take into account malware classification, which requires methods that determine the family of malware rather than only stating its maliciousness.
However, while missing, this inclusion would require the adaptation of metrics to this domain, taking into account attribution mistakes (i.e., determine the wrong malware family) on the present and future data, while also re-defining robustness.
The latter would require rethinking the threat model, which now focuses only on evasion rather than wrong attribution.
Hence, including such a task in our benchmark would have required a different approach, also regarding the models proposed in the state of the art, which are mostly tailored to detection rather than classification.

\section{Future Work and Conclusions}
\mypar{Future Work.} We highlight possible extensions of our methodology that could also inspire further research activities.

\mysubpar{Classification over time.} While not only we want to expand our methodology to address malware classification as well, we are also willing to investigate how to quantify this task over time, taking into account the change in the family distributions on different bins.
This would also fairly characterize how models perform in the presence of new families appearing at some point in time, while others disappear, thus being more difficult to observe in the future.

\mysubpar{Benchmarking more detectors.} 
We are working to expand the results of the benchmark towards more models, also including \emph{dynamic analysis detectors}, i.e., models trained on reports obtained by executing or emulating programs into isolated environments.
While this could require investigating whether other metrics can be used, the current version of \exebench can already include all models under the same leaderboard, even if different pre-processing is applied to the programs.

\mysubpar{Diversify adversarial attacks.} While the metric we have built is general enough to encompass the presence of, in theory, all possible attacks, we will expand the results of the benchmark by including a more comprehensive set of previously-proposed strategies~\cite{lucas2021malware, song2022mab, kozak2024creating}.
These would grant \exebench a pivotal impact on this research topic, standing as a fair methodology to assess robustness of detectors.

\mypar{Conclusions.}
We propose \exebench, a novel and systematic evaluation procedure to rank Windows malware detectors w.r.t. their predictive capabilities on the present and the future data, robustness, and computational requirements.
This is achieved through the definition of a metric that takes into account these quantities, thus creating a general leaderboard of models.
Through our benchmark, we pinpoint the need for evaluations considering multiple axes, since performance on test is, alone, useless to understand the aspects needed in production environments, like stability in time and robustness.
From our results, feature engineering seems to be a winning strategy to balance all these requirements, while some methodologies (unsuitable under the evaluated criteria) should be avoided due to their utter inefficacy.
We believe that \exebench~can provide a starting point for practitioners to pick up a model to deploy depending on the metrics deemed more relevant to them, while also standing as a systematic evaluation procedure to rank all the future models that will be released in the future.

\section*{Acknowledgment}
This work has been partially supported by FISA-2023-00128 funded by the MUR program “Fondo italiano per le scienze applicate”; and by PNRR MUR Project SERICS (PE00000014) and "Future Artificial Intelligence Research (FAIR)", funded by the European Union – NextGenerationEU, PE00000013 CUP J33C24000420007.
D.~Gibert was supported by grant RYC2023-043607-I funded by MICIU/AEI/10.13039/501100011033 and FSE+.
M.~Kozak was supported by MEYS of the Czech Republic, grant No. SGS26/187/OHK3/3T/18 of the Grant Agency, Czech Technical University in Prague.
D.~Trizna was enrolled in the Italian National Doctorate on Artificial Intelligence run by Sapienza University of Rome in collaboration with the University of Genova. 
This research project was made possible through the access granted by the Galician Supercomputing Center (CESGA) to its supercomputing infrastructure. 
The supercomputer FinisTerrae III and its permanent data storage system have been funded by the NextGeneration EU 2021 Recovery, Transformation and Resilience Plan, ICT2021-006904, and also from the Pluriregional Operational Programme of Spain 2014-2020 of the European Regional Development Fund (ERDF), ICTS-2019-02-CESGA-3, and from the State Programme for the Promotion of Scientific and Technical Research of Excellence of the State Plan for Scientific and Technical Research and Innovation 2013-2016 State subprogramme for scientific and technical infrastructures and equipment of ERDF, CESG15-DE-3114.

\bibliographystyle{IEEEtran}
\bibliography{biblio}

@article{demetrio2021adversarial,
    author = {Demetrio, Luca and Coull, Scott E. and Biggio, Battista and Lagorio, Giovanni and Armando, Alessandro and Roli, Fabio},
    title = {Adversarial EXEmples: A Survey and Experimental Evaluation of Practical Attacks on Machine Learning for Windows Malware Detection},
    year = {2021},
    
    volume = {24},
    number = {4},
    issn = {2471-2566},
    doi = {10.1145/3473039},
    journal = {ACM Trans. Priv. Secur.},
    month = sep,
    articleno = {27},
    numpages = {31},
}

@Article{imran2024evaluating,
    AUTHOR = {Imran, Muhammad and Appice, Annalisa and Malerba, Donato},
    TITLE = {Evaluating Realistic Adversarial Attacks against Machine Learning Models for Windows PE Malware Detection},
    JOURNAL = {Future Internet},
    VOLUME = {16},
    YEAR = {2024},
    NUMBER = {5},
    ARTICLE-NUMBER = {168},
    ISSN = {1999-5903},
    DOI = {10.3390/fi16050168}
}

@article{louthanova2024comparison,
    author = {Louth{\'a}nov{\'a}, Pavla and Koz{\'a}k, Matou{\v s} and Jure{\v c}ek, Martin and Stamp, Mark and Di Troia, Fabio},
    date = {2024/11/01},
    doi = {10.1007/s11416-024-00519-z},
    id = {Louth{\'a}nov{\'a}2024},
    isbn = {2263-8733},
    journal = {Journal of Computer Virology and Hacking Techniques},
    number = {4},
    pages = {623--639},
    title = {A comparison of adversarial malware generators},
    volume = {20},
    year = {2024}
}

@inproceedings{verwer2020robust,
    author = {Verwer, Sicco and Nadeem, Azqa and Hammerschmidt, Christian and Bliek, Laurens and Al-Dujaili, Abdullah and O'Reilly, Una-May},
    title = {The Robust Malware Detection Challenge and Greedy Random Accelerated Multi-Bit Search},
    year = {2020},
    isbn = {9781450380942},
    
    doi = {10.1145/3411508.3421374},
    booktitle = {Proc. of 13th ACM Workshop on Art. Int. and Security},
    pages = {61–70},
    numpages = {10},
}

@INPROCEEDINGS{fahim2025optimized,
  author={Fahim, Abrar and Dey, Shamik and Absur, Md. Nurul and Kamrul Siam, Md and Huque, Md. Tahmidul and Jafor Godhuli, Jafreen},
  booktitle={IEEE 14th Int. Conf. on Communication Systems and Network Technologies}, 
  title={Optimized Approaches to Malware Detection: A Study of Machine Learning and Deep Learning Techniques}, 
  year={2025},
  volume={},
  number={},
  pages={269-275},
  doi={10.1109/CSNT64827.2025.10968061}
}

@ARTICLE{vinayakumar2019robust,
  author={Vinayakumar, R. and Alazab, Mamoun and Soman, K. P. and Poornachandran, Prabaharan and Venkatraman, Sitalakshmi},
  journal={IEEE Access}, 
  title={Robust Intelligent Malware Detection Using Deep Learning}, 
  year={2019},
  volume={7},
  number={},
  pages={46717-46738},
  doi={10.1109/ACCESS.2019.2906934}
}

@ARTICLE{2018arXiv180404637A,
  author = {{Anderson}, H.~S. and {Roth}, P.},
  title = "{EMBER: An Open Dataset for Training Static PE Malware Machine Learning Models}",
  journal = {ArXiv e-prints},
  archivePrefix = "arXiv",
  eprint = {1804.04637},
  primaryClass = "cs.CR",
  year = 2018,
  month = apr,
  adsurl = {http://adsabs.harvard.edu/abs/2018arXiv180404637A},
}

@inproceedings{DBLP:conf/aaai/RaffBSBCN18,
  author       = {Edward Raff and
                  Jon Barker and
                  Jared Sylvester and
                  Robert Brandon and
                  Bryan Catanzaro and
                  Charles K. Nicholas},
  title        = {Malware Detection by Eating a Whole {EXE}},
  booktitle    = {Workshop of the The 32nd {AAAI} Conference on Art. Int.},
  year         = {2018},
}

@inproceedings{iclr_avastconv,
  author       = {Marek Krčál and
                  Ondřej Švec and 
                  Martin Bálek and
                  Otakar Jašek},
  title        = {Deep Convolutional Malware Classifiers Can Learn from Raw Executables and Labels Only},
  booktitle    = {ICLR 2018 Workshop},
  year         = {2018},
}

@INPROCEEDINGS{8844623,
  author={Coull, Scott E. and Gardner, Christopher},
  booktitle={IEEE Security and Privacy Workshops (SPW)}, 
  title={Activation Analysis of a Byte-Based Deep Neural Network for Malware Classification}, 
  year={2019},}

@article{GIBERT2021102159,
title = {Auditing static machine learning anti-Malware tools against metamorphic attacks},
journal = {Computers \& Security},
volume = {102},
pages = {102159},
year = {2021},
doi = {https://doi.org/10.1016/j.cose.2020.102159},
author = {Daniel Gibert and Carles Mateu and Jordi Planes and Joao Marques-Silva},
}

@inproceedings{DBLP:conf/cvpr/HeZRS16,
  author       = {Kaiming He and
                  Xiangyu Zhang and
                  Shaoqing Ren and
                  Jian Sun},
  title        = {Deep Residual Learning for Image Recognition},
  booktitle    = {{IEEE} Conference on Computer Vision and Pattern Recognition},
  pages        = {770--778},
  year         = {2016},
  doi          = {10.1109/CVPR.2016.90},
}

@inproceedings{10.1145/2016904.2016908,
  author       = {Lakshmanan Nataraj and
                  S. Karthikeyan and
                  G. Jacob and
                  B. S. Manjunath},
  title        = {Malware images: visualization and automatic classification},
  booktitle    = {8th Int. Symposium on Visualization for Cyber Security (VizSec)},
  pages        = {4},
  publisher    = {{ACM}},
  year         = {2011},
  doi          = {10.1145/2016904.2016908},
}

@article{DBLP:journals/virology/GibertMPV19,
  author       = {Daniel Gibert and
                  Carles Mateu and
                  Jordi Planes and
                  Ramon Vicens},
  title        = {Using convolutional neural networks for classification of malware
                  represented as images},
  journal      = {J. Comput. Virol. Hacking Tech.},
  volume       = {15},
  number       = {1},
  pages        = {15--28},
  year         = {2019},
  doi          = {10.1007/S11416-018-0323-0},
}

@inproceedings{gibert2023_randomizedsmoothing,
	booktitle = {28th Eu. Sym. on Research in Computer Security Workshops},      
	title={Towards a Practical Defense against Adversarial Attacks on Deep Learning-based Malware Detectors via Randomized Smoothing}, 
	author={Daniel Gibert and Giulio Zizzo and Quan Le},
	publisher    = {Springer},
	year={2023}
}

@inproceedings{huang2023rsdel,
	author    = {Huang, Zhuoqun and Marchant, Neil and Lucas, Keane and Bauer, Lujo and Ohrimenko, Olya and Rubinstein, Benjamin I. P.},
	title     = {{RS-Del}: Edit Distance Robustness Certificates for Sequence Classifiers via Randomized Deletion},
	year      = {2023},
	booktitle = {Adv. in Neural Information Processing Systems},
}

@inproceedings{10.1145/3605764.3623914,
author = {Gibert, Daniel and Zizzo, Giulio and Le, Quan},
title = {Certified Robustness of Static Deep Learning-based Malware Detectors against Patch and Append Attacks},
year = {2023},
doi = {10.1145/3605764.3623914},
booktitle = {Proc. of 16th ACM Workshop on Art. Int. and Security},
pages = {173–184},
numpages = {12},
}

@article{saha2024drsm,
  title={Drsm: De-randomized smoothing on malware classifier providing certified robustness},
  author={Saha, S and Wang, W and Kaya, Y and Feizi, S and Dumitras, T},
  journal={ICLR 2024},
  year={2024}
}

@ARTICLE{10506708,
  author={Gibert, Daniel and Zizzo, Giulio and Le, Quan and Planes, Jordi},
  journal={IEEE Access}, 
  title={Adversarial Robustness of Deep Learning-Based Malware Detectors via (De)Randomized Smoothing}, 
  year={2024},
  volume={12},
  number={},
  pages={61152-61162},
  doi={10.1109/ACCESS.2024.3392391}}

@misc{gibert2024certifiedadversarialrobustnessmachine,
      title={Certified Adversarial Robustness of Machine Learning-based Malware Detectors via (De)Randomized Smoothing}, 
      author={Daniel Gibert and Luca Demetrio and Giulio Zizzo and Quan Le and Jordi Planes and Battista Biggio},
      year={2024},
      eprint={2405.00392},
      archivePrefix={arXiv},
      primaryClass={cs.CR},
      url={https://arxiv.org/abs/2405.00392}, 
}

@article{ke2017lightgbm,
  title={Lightgbm: A highly efficient gradient boosting decision tree},
  author={Ke, Guolin and Meng, Qi and Finley, Thomas and Wang, Taifeng and Chen, Wei and Ma, Weidong and Ye, Qiwei and Liu, Tie-Yan},
  journal={Adv. in Neural Information Processing Systems},
  volume={30},
  year={2017}
}

@inproceedings{kolosnjaji2018adversarial,
  title={Adversarial malware binaries: Evading deep learning for malware detection in executables},
  author={Kolosnjaji, Bojan and Demontis, Ambra and Biggio, Battista and Maiorca, Davide and Giacinto, Giorgio and Eckert, Claudia and Roli, Fabio},
  booktitle={26th European signal processing conference (EUSIPCO)},
  pages={533--537},
  year={2018},
  organization={IEEE}
}

@inproceedings{lucas2021malware,
  title={Malware makeover: Breaking ml-based static analysis by modifying executable bytes},
  author={Lucas, Keane and Sharif, Mahmood and Bauer, Lujo and Reiter, Michael K and Shintre, Saurabh},
  booktitle={Proc. of ACM Asia Conference on Computer and Communications Security},
  pages={744--758},
  year={2021}
}

@inproceedings{pierazzi2020intriguing,
  title={Intriguing properties of adversarial ml attacks in the problem space},
  author={Pierazzi, Fabio and Pendlebury, Feargus and Cortellazzi, Jacopo and Cavallaro, Lorenzo},
  booktitle={IEEE symposium on security and privacy (SP)},
  pages={1332--1349},
  year={2020},
  organization={IEEE}
}

@inproceedings{trizna2022quo,
  title={Quo Vadis: hybrid machine learning meta-model based on contextual and behavioral malware representations},
  author={Trizna, Dmitrijs},
  booktitle={Proc. of 15th ACM Workshop on Art. Int. and Security},
  pages={127--136},
  year={2022}
}

@inproceedings{pendlebury2019tesseract,
  title={$\{$TESSERACT$\}$: Eliminating experimental bias in malware classification across space and time},
  author={Pendlebury, Feargus and Pierazzi, Fabio and Jordaney, Roberto and Kinder, Johannes and Cavallaro, Lorenzo},
  booktitle={28th {USENIX} sec. sym.},
  pages={729--746},
  year={2019}
}

@inproceedings{cina2025attackbench,
  title={Attackbench: Evaluating gradient-based attacks for adversarial examples},
  author={Cin{\`a}, Antonio Emanuele and Rony, J{\'e}r{\^o}me and Pintor, Maura and Demetrio, Luca and Demontis, Ambra and Biggio, Battista and Ayed, Ismail Ben and Roli, Fabio},
  booktitle={Proc. of AAAI Conference on Art. Int.},
  volume={39},
  number={3},
  pages={2600--2608},
  year={2025}
}

@article{pintor2021fast,
  title={Fast minimum-norm adversarial attacks through adaptive norm constraints},
  author={Pintor, Maura and Roli, Fabio and Brendel, Wieland and Biggio, Battista},
  journal={Adv. in Neural Information Processing Systems},
  volume={34},
  pages={20052--20062},
  year={2021}
}

@article{biggio2018wild,
  title={Wild patterns: Ten years after the rise of adversarial machine learning},
  author={Biggio, Battista and Roli, Fabio},
  journal={Elsevier Pattern Recognition},
  volume={84},
  pages={317--331},
  year={2018},
}

@article{demetrio2021functionality,
  title={Functionality-preserving black-box optimization of adversarial windows malware},
  author={Demetrio, Luca and Biggio, Battista and Lagorio, Giovanni and Roli, Fabio and Armando, Alessandro},
  journal={IEEE Tran. on Information Forensics and Security},
  volume={16},
  pages={3469--3478},
  year={2021},
}

@inproceedings{10.1145/2857705.2857713,
author = {Ahmadi, Mansour and Ulyanov, Dmitry and Semenov, Stanislav and Trofimov, Mikhail and Giacinto, Giorgio},
title = {Novel Feature Extraction, Selection and Fusion for Effective Malware Family Classification},
year = {2016},
isbn = {9781450339353},
booktitle = {Proc. of 6th ACM Conference on Data and Application Security and Privacy},
pages = {183–194},
numpages = {12},
}

@INPROCEEDINGS{7413680,
  author={Saxe, Joshua and Berlin, Konstantin},
  booktitle={10th Int. Conf. on Malicious and Unwanted Software (MALWARE)}, 
  title={Deep neural network based malware detection using two dimensional binary program features}, 
  year={2015},
  volume={},
  number={},
  pages={11-20}
}

@article{gibert2020rise,
  title={The rise of machine learning for detection and classification of malware: Research developments, trends and challenges},
  author={Gibert, Daniel and Mateu, Carles and Planes, Jordi},
  journal={Elsevier Journal of Network and Computer Applications},
  volume={153},
  pages={102526},
  year={2020},
}

@inproceedings{lucas2023adversarial,
  title={Adversarial training for $\{$Raw-Binary$\}$ malware classifiers},
  author={Lucas, Keane and Pai, Samruddhi and Lin, Weiran and Bauer, Lujo and Reiter, Michael K and Sharif, Mahmood},
  booktitle={32nd {USENIX} sec. sym.},
  pages={1163--1180},
  year={2023}
}

@article{kozak2025updating,
  title={Updating Windows malware detectors: Balancing robustness and regression against adversarial EXEmples},
  author={Kozak, Matous and Demetrio, Luca and Trizna, Dmitrijs and Roli, Fabio},
  journal={Elsevier Computers \& Security},
  volume={155},
  pages={104466},
  year={2025},
}

@article{kozak2024creating,
  title={Creating valid adversarial examples of malware},
  author={Koz{\'a}k, Matou{\v{s}} and Jure{\v{c}}ek, Martin and Stamp, Mark and Troia, Fabio Di},
  journal={Springer Journal of Computer Virology and Hacking Techniques},
  volume={20},
  number={4},
  pages={607--621},
  year={2024},
}

@inproceedings{song2022mab,
  title={Mab-malware: A reinforcement learning framework for blackbox generation of adversarial malware},
  author={Song, Wei and Li, Xuezixiang and Afroz, Sadia and Garg, Deepali and Kuznetsov, Dmitry and Yin, Heng},
  booktitle={Proc. of ACM on Asia conference on computer and communications security},
  pages={990--1003},
  year={2022}
}

@inproceedings{demontis2019adversarial,
  title={Why do adversarial attacks transfer? explaining transferability of evasion and poisoning attacks},
  author={Demontis, Ambra and Melis, Marco and Pintor, Maura and Jagielski, Matthew and Biggio, Battista and Oprea, Alina and Nita-Rotaru, Cristina and Roli, Fabio},
  booktitle={28th {USENIX} sec. sym.},
  pages={321--338},
  year={2019}
}

@article{friedman2001greedy,
  title={Greedy function approximation: a gradient boosting machine},
  author={Friedman, Jerome H},
  journal={Annals of statistics},
  pages={1189--1232},
  year={2001},
  publisher={JSTOR}
}

@inproceedings{DBLP:conf/icml/CohenRK19,
  author       = {Jeremy Cohen and
                  Elan Rosenfeld and
                  J. Zico Kolter},
  title        = {Certified Adversarial Robustness via Randomized Smoothing},
  booktitle    = {Proc. of 36th Int. Conf. on Machine Learning},
  volume       = {97},
  pages        = {1310--1320},
  year         = {2019},
}

@inproceedings{NEURIPS2020_47ce0875,
 author = {Levine, Alexander and Feizi, Soheil},
 booktitle = {Adv. in Neural Information Processing Systems},
 pages = {6465--6475},
 title = {(De)Randomized Smoothing for Certifiable Defense against Patch Attacks},
 volume = {33},
 year = {2020}
}

@article{paszke2019pytorch,
  title={Pytorch: An imperative style, high-performance deep learning library},
  author={Paszke, Adam and Gross, Sam and Massa, Francisco and Lerer, Adam and Bradbury, James and Chanan, Gregory and Killeen, Trevor and Lin, Zeming and Gimelshein, Natalia and Antiga, Luca and others},
  journal={Adv. in Neural Information Processing Systems},
  volume={32},
  year={2019}
}

@inproceedings{10.1145/3658644.3690208,
author = {Lucas, Keane and Lin, Weiran and Bauer, Lujo and Reiter, Michael K. and Sharif, Mahmood},
title = {Training Robust ML-based Raw-Binary Malware Detectors in Hours, not Months},
year = {2024},
isbn = {9798400706363},
booktitle = {Proc. of 2024 on ACM SIGSAC Conference on Computer and Communications Security},
pages = {124–138},
numpages = {15},
}

@inproceedings{croce2021robustbench,
  title     = {RobustBench: a standardized adversarial robustness benchmark},
  author    = {Croce, Francesco and Andriushchenko, Maksym and Sehwag, Vikash and Debenedetti, Edoardo and Flammarion, Nicolas and Chiang, Mung and Mittal, Prateek and Matthias Hein},
  booktitle = {35th Conf. on Neural Information Processing Systems Datasets and Benchmarks Track},
  year      = {2021},
}

@inproceedings{joyce2025ember2024,
  title={EMBER2024-A Benchmark Dataset for Holistic Evaluation of Malware Classifiers},
  author={Joyce, Robert J and Miller, Gideon and Roth, Phil and Zak, Richard and Zaresky-Williams, Elliott and Anderson, Hyrum and Raff, Edward and Holt, James},
  booktitle={Proc. of 31st ACM SIGKDD Conf. on Knowledge Discovery and Data Mining V. 2},
  pages={5516--5526},
  year={2025}
}

@article{harang2020sorel,
  title={SOREL-20M: A large scale benchmark dataset for malicious PE detection},
  author={Harang, Richard and Rudd, Ethan M},
  journal={arXiv preprint},
  year={2020}
}

@inproceedings{kurlandski2026,
  title={Beyond Raw Bytes: Towards Large Language Models},
  author={Kurlandski, Luke and Berger, Harel and Pan, Yin and Wright, Matthew},
  booktitle={Proc. of 33rd Network and Distributed Systems (NDSS) Symposium},
  year={2026}
}

@ARTICLE{10145856,
  author={Chen, Yi-Hsien and Lin, Si-Chen and Huang, Szu-Chun and Lei, Chin-Laung and Huang, Chun-Ying},
  journal={IEEE Transactions on Information Forensics and Security}, 
  title={Guided Malware Sample Analysis Based on Graph Neural Networks}, 
  year={2023},
  volume={18},
  number={},
  pages={4128-4143},
  doi={10.1109/TIFS.2023.3283913}}

\appendix
\section{Appendix}

\mypar{A: Area Under the Curve fallacy}
\label{appendix:auc}    
The Area Under the Curve (AUC) quantifies the area below the curve, empirically approximated with a finite number of rectangles or trapezoids covering such region, without addressing the number of samples used to produce each point in the curve.
Given 2 splits (considered as one single unit away on the x axis), one with 2 samples and F1 score of 0, and one with 100 samples and F1 score of 0.8, the total AUC would be $1 * (0 + 0.8) / 2 = 0.4$.
However, the result is penalized by the first bin being under-represented, weighting the same on the average.
On the contrary, by weighting this number on the amount of samples contained into the bins, we would compute $(0 * \frac{2}{102} ) + 0.8 * \frac{100}{102}) = 0.78$.

\mypar{B: Training Time in Hours}
\label{appendix:training_time}
We report in \autoref{tab:training_time} the estimated training time (in hours) of all models.
Except for one model (i.e., MalConvRS requires $\sim146$ hours of training time), training does not pose a huge bottleneck, since such operation is not a frequent as inference.
On the contrary, while feature extraction seems slow (EMBER GBDT requires roughly 7 hours to process and train 600k programs), we remark that samples are only processed once, thus speeding up next updates (training alone on 600k featurized samples only requires 4 minutes), differently from deep neural networks that must be re-trained from scratch.
\begin{table}[h]
\centering
\footnotesize
\begin{tabular}{lc|lc}
\toprule
\textbf{Model} & \textbf{Time (h)} &
\textbf{Model} & \textbf{Time (h)} \\
\midrule
AvastConvRsDel & 0.46 & AvastConvSDRS & 2.95 \\
BBDnnRsDel & 0.83 & AvastConvKDRS & 3.54 \\
MalConvSDRS & 1.04 & BBDnnFDRS & 4.99 \\
MalConvRDRS & 1.04 & NGramConvKDRS & 5.18 \\
NGramConvRsDel & 1.32 & BBDnnKDRS & 5.29 \\
BBDnnSDRS & 1.47 & BBDnn & 6.00 \\
BBDnnRDRS & 1.47 & MalConvKDRS & 6.44 \\
AvastConvFDRS & 2.02 & EmberGBDT & 7.18 \\
NGramConvRDRS & 2.05 & AvastConv & 7.62 \\
NGramConvSDRS & 2.05 & AvastConvRS & 7.99 \\
MalConvFDRS & 2.13 & BBDnnRS & 10.51 \\
ResNet18 & 2.19 & NGramConvRS & 18.12 \\
NGramConvFDRS & 2.44 & MalConv & 19.54 \\
MalConvRsDel & 2.91 & NGramConv & 23.96 \\
AvastConvRDRS & 2.95 & MalConvRS & 146.47 \\
\bottomrule
\end{tabular}
\caption{Training time (hours) of all the considered models.}
\label{tab:training_time}
\end{table}

\textcolor{black}{
\mypar{C: Metadata gathered from VirusTotal}}
\begin{table}[h]
\setlength{\tabcolsep}{2pt}
\resizebox{\linewidth}{!}{
\begin{tabular}{@{}cccccccc@{}}
\toprule
\multirow{2}{*}{\textbf{Bin}} & \multirow{2}{*}{\textbf{Tot.}} & \multicolumn{3}{c}{\textbf{First Submission}} & \multicolumn{3}{c}{\textbf{First Seen}} \\ \cmidrule(l){3-8} 
 &  & \textbf{Total} & \textbf{Train} & \textbf{Test} & \textbf{Total} & \textbf{Train} & \textbf{Test} \\ \midrule
2019-Q1 & \multicolumn{1}{c|}{2,440} & 1846 (75.7\%) & 1035 (56.1\%) & \multicolumn{1}{c|}{811 (43.9\%)} & 105 (4.3\%) & 53 (50.5\%) & 52 (49.5\%) \\
2019-Q2 & \multicolumn{1}{c|}{1,895} & 1618 (85.4\%) & 1356 (83.8\%) & \multicolumn{1}{c|}{262 (16.2\%)} & 11 (0.6\%) & 11 (100.0\%) & 0 (0.0\%) \\
2019-Q3 & \multicolumn{1}{c|}{1,043} & 262 (25.1\%) & 223 (85.1\%) & \multicolumn{1}{c|}{39 (14.9\%)} & 47 (4.5\%) & 40 (85.1\%) & 7 (14.9\%) \\
2020-Q1 & \multicolumn{1}{c|}{2,500} & 508 (20.3\%) & 336 (66.1\%) & \multicolumn{1}{c|}{172 (33.9\%)} & 37 (1.5\%) & 26 (70.3\%) & 11 (29.7\%) \\
2020-Q2 & \multicolumn{1}{c|}{1,401} & 707 (50.5\%) & 502 (71.0\%) & \multicolumn{1}{c|}{205 (29.0\%)} & 95 (6.8\%) & 56 (58.9\%) & 39 (41.1\%) \\
2020-Q3 & \multicolumn{1}{c|}{2,198} & 1052 (47.9\%) & 850 (80.8\%) & \multicolumn{1}{c|}{202 (19.2\%)} & 211 (9.6\%) & 113 (53.6\%) & 98 (46.4\%) \\
2021-Q1 & \multicolumn{1}{c|}{4,128} & 1249 (30.3\%) & 996 (79.7\%) & \multicolumn{1}{c|}{253 (20.3\%)} & 221 (5.4\%) & 96 (43.4\%) & 125 (56.6\%) \\
2021-Q2 & \multicolumn{1}{c|}{3,494} & 1528 (43.7\%) & 1116 (73.0\%) & \multicolumn{1}{c|}{412 (27.0\%)} & 287 (8.2\%) & 144 (50.2\%) & 143 (49.8\%) \\
2021-Q3 & \multicolumn{1}{c|}{5,541} & 1796 (32.4\%) & 1169 (65.1\%) & \multicolumn{1}{c|}{627 (34.9\%)} & 214 (3.9\%) & 115 (53.7\%) & 99 (46.3\%) \\
2022-Q1 & \multicolumn{1}{c|}{18,864} & 3571 (18.9\%) & 1848 (51.8\%) & \multicolumn{1}{c|}{1723 (48.2\%)} & 1078 (5.7\%) & 648 (60.1\%) & 430 (39.9\%) \\ \bottomrule \addlinespace 
\multicolumn{2}{@{}l}{$\geq$ Collection Date} & 6{,}344 (44.9\%) & 5{,}936 & \multicolumn{1}{c|}{408} & - & - & - \\ \bottomrule
\end{tabular}}
\caption{Summary of metadata collected from VirusTotal.
The last row considers only samples first submitted after the collection date (Train $\geq$ 2022-02, Test $\geq$ 2022-05).}
\label{tab:vt_timestamp_analysis}
\end{table}

\textcolor{black}{
\mypar{D: Rank Correlation Analysis}}

\begin{table}[h]
\centering
\setlength{\tabcolsep}{3pt}
\resizebox{0.7\linewidth}{!}{
\begin{tabular}{lcc}
\toprule
& \multicolumn{2}{c}{\textbf{Leave-One-Out}} \\
\cmidrule(lr){2-3}
\textbf{Metric} & Pearson $r$ ($p$-val.) & Spearman $\rho$ ($p$-val.) \\
\midrule
Performance  & +0.5291 (0.0026) & +0.3419 (0.0644) \\
Temporal     & +0.3984 (0.0292) & +0.5622 (0.0012) \\
Robustness   & \textbf{+0.2162} (0.2513) & +0.3673 (0.0459) \\
Inference    & +0.2623 (0.1614) & \textbf{+0.2538} (0.1759) \\
\bottomrule
\end{tabular}}
\caption{Correlation with final Rank (Leave-One-Out).}
\label{tab:correlation}
\end{table}

\end{document}